\documentclass[nofootinbib,aps,pra,twocolumn,english]{revtex4-1}
\pdfoutput=1

\usepackage{orcidlink}

\usepackage{graphicx}
\usepackage{wrapfig,enumerate,slashed}
\usepackage[utf8]{inputenc}
\usepackage{adjustbox}
\usepackage{color, colortbl}
\usepackage{tabularray}

\usepackage[english]{babel}
\usepackage[T1]{fontenc}
\usepackage{amsmath}

\usepackage{float}

\usepackage{appendix}
\usepackage{placeins}

\usepackage{amsfonts}
\usepackage{amsmath}
\usepackage{wasysym} 
\usepackage{graphicx}
\usepackage{comment}
\usepackage{slashed}
\usepackage{booktabs}
\usepackage{listings}
\renewcommand{\d}{{\mathrm{d}}}
\newcommand{\be}{\begin{eqnarray}}
\newcommand{\ee}{\end{eqnarray}}

\newcommand{\bee}{\begin{eqnarray}}
\newcommand{\eee}{\end{eqnarray}}
\newcommand{\beeq}{\begin{equation}}
\newcommand{\eeeq}{\end{equation}}

\usepackage{listings}
\usepackage{color}
\usepackage{hyperref}
\usepackage{tabularray}

\definecolor{dkgreen}{rgb}{0,0.6,0}
\definecolor{gray}{rgb}{0.5,0.5,0.5}
\definecolor{mauve}{rgb}{0.58,0,0.82}
\definecolor{cyan}{rgb}{0.88,1,1}
\definecolor{TopRow}{rgb}{0.4,0.7,1}
\definecolor{lblue}{rgb}{0.8,0.9,1}

\makeatletter

    \def\CT@@do@color{%
      \global\let\CT@do@color\relax
            \@tempdima\wd\z@
            \advance\@tempdima\@tempdimb
            \advance\@tempdima\@tempdimc
    \advance\@tempdimb\tabcolsep
    \advance\@tempdimc\tabcolsep
    \advance\@tempdima2\tabcolsep
            \kern-\@tempdimb
            \leaders\vrule
                    \hskip\@tempdima\@plus  1fill
            \kern-\@tempdimc
            \hskip-\wd\z@ \@plus -1fill }
    \makeatother
\begin{document}
\title{Ising Machines for Diophantine Problems in Physics}
%\preprint{}

\author{Steven~A.~Abel \orcidlink{0000-0003-1213-907X} }
\email{steve.abel@durham.ac.uk }
\author{Luca~A.~Nutricati \orcidlink{0000-0002-5045-5113}}
\email{luca.a.nutricati@durham.ac.uk}

\affiliation{\vspace{0.1cm} Institute for Particle Physics Phenomenology, Durham University, Durham DH1 3LE, UK}
\affiliation{Department of Mathematical Sciences, Durham University, Durham DH1 3LE, UK}

\begin{abstract}
{\small
Diophantine problems arise frequently in physics, in for example anomaly cancellation conditions, string consistency conditions and so forth. We present methods to solve such problems to high order on annealers that are based on the quadratic Ising Model. This is the intrinsic framework for both quantum annealing and for common forms of  classical simulated annealing. We demonstrate the method on so-called Taxicab numbers (discovering some apparently new ones), and on the realistic problem of anomaly cancellation in $U(1)$ extensions of the Standard Model.    }
\end{abstract}

\maketitle

% make two column:

\flushbottom

%%%%%%%%%%%%%%%%%%%%%%%%%%%%%%%%%%%%%%%%%%%%%%%

\section{\label{Sec:Intro}Introduction}

As well as being of intrinsic interest in number theory, Diophantine problems play an important role in fundamental physics. The purpose of this paper is to develop  methods for solving the kinds of Diophantine problems that frequently occur in particle physics, where for example they appear in anomaly cancellation conditions as systems of cubic equations. They also
appear  in index theorems, in consistency conditions in string theory, in problems relevant for computing the effective potential in string theory and for finding vacua with small cosmological constant, as well as in powerful nonperturbative methods that are used in field theory, such as 't\,Hooft anomaly matching~\cite{tHooft:1979rat}, and a multitude of other applications. 

Unfortunately Diophantine problems are also notoriously difficult to solve. Sometimes they can be significantly simplified using for example Gr\"obner basis methods, but typically such a problem is computationally hard. This is certainly the case for the typical anomaly cancellation problem, which entails the solving of a coupled set of cubic equations with an independent rational variable appearing for every charge of every particle. In a system whose size is comparable to that of say the Standard Model of particle physics, an exhaustive scan becomes infeasible even when the domain of allowed charges is restricted. (See for example Refs.~\cite{Allanach:2018vjg,Allanach:2019uuu,Allanach:2020zna,Allanach:2021bfe}.) 
Indeed determining precisely which complexity class a problem falls into is itself an important question. (See for example Ref.~\cite{TUNG1987324}, and for discussions in the string theory context  Refs.~\cite{Denef:2006ad,Denef:2017cxt,Halverson:2018cio,Halverson:2019vmd}.)

For such problems, scaleable Ising hardware solvers, which form the basis of both simulated and quantum annealers (as introduced in Ref.~\cite{PhysRevE.58.5355,
doi:10.1126/science.1057726}),  could have huge impact, particularly as NP problems can be formulated as Ising problems with only polynomial overhead \cite{RevModPhys.80.1061,10.3389/fphy.2014.00005}. (For some practical applications of Ising machines and for recent reviews see Refs.~\cite{Hauke_2020,Mohseni}). 

This paper will show how to encode Diophantine problems of the kind described above onto such machines. We shall do this by focusing on two Diophantine tasks. 

The first is the purely number theoretic one of finding what we will refer to generally as ``Taxicab'' numbers, namely those numbers that can be expressed in more than one way as sums of equal powers. The most famous example is the number of Hardy and Ramanujan's eponymous taxi, ${\rm Ta}(2)=1729$. This is the smallest of the following list of numbers, all of which are expressible as the sum of two cubes in two different ways: 
\begin{align}
&1729 ~=~ 1^3 + 12^3 ~=~ 9^3 + 10^3, \nonumber\\
&4104 ~=~ 9^3 + 15^3 ~=~ 16^3 + 2^3, \nonumber \\
&20683 ~=~  24^3 + 19^3 ~=~ 10^3 + 27^3, \nonumber\\
&32832 ~=~ 32^3 + 4^3 ~=~ 18^3 + 30^3, \nonumber\\
&...
\label{eq:taxi_solutions}
\end{align}
We shall use the notation $(k,m,n)$, to refer to such numbers, where $k$ is the power, while $m$ and $n$ are the number of terms on each side. Thus Ta$(2) ~=~ 1729$ is defined to be the smallest $(3,2,2)$ number, while Fermat's theorem is the statement that $(k,1,2)$ numbers only exist for $k=2$. Here we will develop annealing methods  to determine the above list of $(3,2,2)$ numbers (where we consider all numbers in the list to be of interest not just the smallest). We also test our methods on several variations, namely $(4,3,3)$, $(3,1,5)$, $(3,1,7)$, $(3,6,6)$, $(3,7,7)$, $(3,8,8)$. Examples of most of these are known, and can be found in Refs.~\cite{Guy:1994a,Meyrignac,Eulernet,Weisstein,piezas}, although some were discovered only with the advent of high performance computing and appeared relatively recently. However some, such as $(3,7,7)$ and $(3,8,8)$ numbers, do not seem to have been known before. (Indeed as we shall see the latter represent solutions in a search space of size $\sim 10^{24}$.)

 The second task that we will consider is the physical one of finding solutions to the anomaly cancellation conditions of a typical extension of the Standard Model of particle physics. In four space-time dimensions this is as mentioned also a cubic (i.e. homogeneous, third order) Diophantine problem, in which the integers correspond to the numerators of rational gauge charges. (In $2d$ dimensions the  equations are instead order $d+1$). 
 In this case as well, we will find that  a sufficiently  well-crafted encoding  onto an annealer allows one  to solve for anomaly cancellation in the systems considered in Refs.~\cite{Allanach:2019uuu,Allanach:2020zna,Allanach:2021bfe}, at an already relatively advanced level, and in a very short time, and certainly without having to perform any kind of exhaustive scan.

Although we will not consider 't\,Hooft anomaly matching explicitly in our discussion, it is worth mentioning that this nonperturbative procedure is morally a generalisation of the Ta$(2)$ Taxicab problem. That is, when two particle theories are duals of each other, then by 't\,Hooft's argument they must realise the same set of global anomalies in two  different ways. Similarly physical trialities and quadralities are generalisations of the Ta$(3)$ and Ta$(4)$ Taxicab problems (which realise the same number in 3 and 4 different ways respectively). Such systems exist but are comparatively rare \cite{Gadde:2013lxa,Franco:2016nwv,Franco:2016tcm}.\\
 
Constructing suitably efficient encodings for these sorts of problems requires significant advancement. Firstly as we shall review we are interested in the kinds of set-ups found in quantum annealers, in which problems are encoded in the Hamiltonian of a quadratic Ising model, which must then be minimised to solve the problem. Thus the crux of the matter in encoding a non-trivial system of cubic- and higher-order Diophantine equations is to implement a reduction procedure that can represent the complete system as a single loss-function represented by a spin-Hamiltonian that is at most quadratic. Of course  Diophantine problems, in particular factorisation, have been considered on Ising model annealers before  \cite{dattani2014quantum,tanburn:2015a,Jiang:2018b,peng2019factoring,warren2019factoring,wang2020prime}, along with quadratic systems of polynomial equations \cite{chang2019quantum, chang2019least}. However both these problems can be mapped into the optimisation of an order four Hamiltonian which can in turn be reduced to a quadratic Ising model suitable for a  quantum annealer, with only two layers of reduction. By contrast, here we will be considering problems of order much higher than two. To accomplish this, we use a procedure to automate the reduction of an arbitrary order Hamiltonian to a quadratic one. This procedure, which iterates that first appearing in Ref.~\cite{rosenberg1975reduction} and then more recently in  Refs.~\cite{tanburn:2015a,dattani:2019a,Hauke_2020,Abel:2022lqr}, is completely problem-independent and therefore potentially applicable to any set of Diophantine equations. It can perform the many layers of reduction required to reach a quadratic spin-Hamiltonian representation of the high order problems we will be considering.

Furthermore, the integers in the Diophantine equations are naturally encoded using a binary representation, which is what 
ultimately will enable the procedure to be much more efficient than a systematic scan. However this in turn leads to large and highly connected Ising model encodings of the Diophantine system of equations. 
Thus to improve efficiency we introduce a technique we refer to as {\it solution-mining}. This innovation begins by finding one solution in the conventional manner. A random perturbation from this solution then serves as the new starting point for the next run. From there the system often tunnels to a new solution nearby, and when it does so this serves as the next starting point, and so on. 
This approach appears to be effective for problems, like anomaly cancellation, that contain many coupled Diophantine equations, and it allows the domain of solutions to wander in the parameter space. We shall describe these techniques in Section~\ref{Sec:Annealer}. 

The rest of the paper is organised as follows. In Section~\ref{Sec:quantumandclassical} we begin with a warm-up problem which is to solve a single Diophantine polynomial equation with two variables. We use this simple problem to discuss how annealing is implemented on a quantum annealer, and to compare the performance of currently available quantum annealers with classical simulated annealers. Our conclusion here is that quantum annealers are only just beginning to catch up with classical simulated annealing in terms of the complexity of problems that can be embedded and solved, but when they do become competitive the potential speed-ups could be substantial for this kind of problem. 
Then Sections~\ref{Sec:TaxiCab} and \ref{Sec:Anomalies} consider the Taxicab problems, and anomalies respectively.

\section{\label{Sec:Annealer}Methods for encoding Diophantine problems}

The general principle behind an Ising machine is to solve  problems by reformulating their solution as the minimisation of a  function $H$ of spin variables $\sigma_\ell ~=~ \pm 1$, where $\ell$ labels the spin sites. On a quantum annealer this  so-called {\it problem-Hamiltonian}, $H$, is forced (by the physical architecture of the annealer) to be quadratic in the spins,
\begin{equation}
\label{eq:isingH}
 H(\sigma_\ell) ~=~ \sum_\ell h_\ell \sigma_\ell + \sum_{\ell m} J_{\ell m} \sigma_\ell \sigma_m~,
\end{equation}
where the spins would of course correspond to physical {qubits}. We can postpone further discussion of the physical realisation of the system (which will be described later) and focus on the embedded abstract problem which will apply to any annealer of this kind. 

Specialising to the  present case, we are interested in solving a set of polynomial  Diophantine equations $f_A(t_i)=0$, where $t_i\in {\mathbb Z} $ are the would-be integer solutions to the problem of interest. These integers will be encoded on the annealer in a binary format, namely  we will use the following encoding:
\beeq
t_i ~=~ \tau_{i,0} + 2 \tau_{i,1} + \dots + 2^{\beta - 1}\tau_{i,\beta-1} + s_i~.
\label{eq:binary_encoding}
\eeeq
We use $\tau$ to denote the binary variables corresponding to a given spin,
\begin{equation}
\label{eq:taufromsig}
    \tau_{i,k} ~=~\frac{1}{2} (1+\sigma_{i,k})~,
\end{equation}
with $\tau_{i,k} \in \{0,1\}$, and where we allow classical integer shifts, $s_i\in\mathbb {Z}$.
These shifts can for example be negative to allow the domain to include negative integers or, as will be the case for solution-mining, they can be adjusted iteratively to explore the search space. 

We would like to use such an encoding to solve the Diophantine equations, and this can {\it in principle} be done by finding the minimum of a loss-function Hamiltonian, 
\begin{equation}
H_{\rm D}~\equiv~\sum_A (f_A(t_i(\sigma_\ell)) )^2~.
\end{equation}
In addition one may wish to add several constraints: for example the Taxicab numbers are usually defined as the {\it smallest} numbers expressible in two different ways. Such conditions can be included with a constraint Hamiltonian, $H_C$, which might in the case of the Taxicab numbers be simply the numbers themselves (since they are positive). Thus we begin with an idealised (i.e. non-quadratic) system, 
\begin{equation}
\label{eq:HdHc}
\tilde{H}(\sigma_\ell)~=~H_D(t_i(\sigma_\ell))+H_C(t_i(\sigma_\ell))~.\end{equation}
Note that solutions to the Diophantine equations all have $H_D=0$, so that constraints imposed by $H_C$ would  independently select the preferred solution. However the converse is generally not true: that is one should  avoid over-weighting the constraints $H_C$ such that competing minima appear that have lower $H_C$ but $H_D\neq 0$. Of course in many cases the desired solutions are very rare, so it is often much more efficient (or more precisely not an NP-hard problem) to simply apply any desired constraints by  post-processing the solutions (e.g. for the Taxicab numbers, one could simply select by hand the smallest number). 

\subsection{Reduction}

\label{subsec:reduction}

For a Diophantine system containing order-$d$ polynomials in $t_i$, the raw Hamiltonian $\tilde{H}$ in Eq.~\eqref{eq:HdHc} is an order-$2d$ polynomial in the spins, $\sigma_\ell$. For example the ${\rm Ta}(2)$ problem yields an order-6 spin-polynomial. Therefore we now arrive at the task of transforming $\tilde{H}$ into an equivalent quadratic problem-Hamiltonian, $H$, that is guaranteed to have the same minima, but which can be implemented on the annealer. 

For this task we shall use the reduction method described in the Appendix of Ref.~\cite{Abel:2022lqr}. This method works by introducing auxilliary spins\footnote{We think it is more accurate to use `auxilliary' to refer to both abstract spins and later to qubits, rather than the quantum computing term, `ancillary'.} to represent pairs of spins in the original Hamiltonian of Eq.~\eqref{eq:HdHc},  and is one of the many methods in the  comprehensive survey of  Ref.~\cite{dattani:2019a}. (We should remark that there exist  ``qubit-saving'' reduction methods that do not require auxilliary spins, but these are more task specific and currently appear to be restricted to reduction of terms in the Hamiltonian with products of 3 or 4 spins~\cite{tanburn:2015a}.)

The method works as follows. We begin with the raw polynomial $\tilde{H}(\sigma_\ell)$ written as a function of binary variables using Eq.~\eqref{eq:taufromsig}. Suppose $\tilde{H}$ has terms involving products of two binary variables $\tau_1$ and $\tau_2$.
Now consider adding to the polynomial $\tilde{H}$ a quadratic term that involves the binary variables together with a new auxiliary variable $\tau_{12}$, which is of the form 
\begin{equation}
Q(\mbox{\small $\tau_{12};\tau_1,\tau_2$}) ~ = ~ \Lambda (\tau_1 \tau_2 - 2 \tau_{12} (\tau_1 + \tau_2) + 3 \tau_{12})~.
 \label{eq:constraint-Hamiltonian}
\end{equation}
Inspection shows that a sufficiently large and positive overall coupling $\Lambda$ enforces $\tau_{12} ~=~\tau_1\tau_2$. Importantly the  minimum at this point has $Q=0$. Therefore we may replace the product $\tau_1\tau_2$ with $\tau_{12}$ wherever it appears within $\tilde{H}$, and the new Hamiltonian is guaranteed to have the same set of minima as the original $\tilde{H}$. Therefore the process can be iterated until one arrives at the desired problem-Hamiltonian which is quadratic in spins, 
and which is schematically of the form 
\begin{align}
H_D+H_C ~&=~ \tilde{H}(\mbox{\small$\tau_1,\tau_2,\ldots, \tau_{12},\tau_{13},\ldots ,\tau_{12,34},\tau_{12,56}\ldots $}) \nonumber \\
& + \sum_{i>j} Q(\mbox{\small$\tau_{ij};\tau_i,\tau_j$}) +\sum_{i<j,k<m}
Q(\mbox{\small$ \tau_{ij,km};\tau_{ij},\tau_{km}$})\nonumber \\
& \qquad ~+\ldots 
\end{align}
with the constraints imposed by the $Q$ terms ensuring that this quadratic Hamiltonian has the same minima as the original order-$2d$ polynomial. 

We can check that the reduction works correctly with the example order-3 Hamiltonian 
\begin{align}
\tilde{H}~&=~ \sigma_1\sigma_2\sigma_3~\nonumber \\
  &\equiv ~ 8 \tau_1\tau_2\tau _3 - 4 \tau_1\tau_2- 4 \tau_1\tau_3 - 4 \tau_2\tau_3 \nonumber \\ 
  & \qquad \qquad \qquad \qquad + 2 \tau_1 +2 \tau_2+ 2 \tau_3~,
\end{align}
where we drop the constant $-1$ in translating to the binary variables. This Hamiltonian has 4 minima at $\sigma_1\sigma_2\sigma_3~=~-1$ (which corresponds in binary language to any one of the $\tau_\ell$ being zero, or all of them), as opposed to the seven solutions to $\tau_1\tau_2\tau _3 ~=~0$. As described above we can reduce the trilinear term by trading $\tau_1 \tau_2$ for an auxiliary binary $\tau_{12}$ and adding the Hamiltonian $Q(\mbox{\small $\tau_{12};\tau_1,\tau_2$}) $. The quadratic  problem-Hamiltonian (in QUBO language) is then 
\begin{align}
H~& = ~ Q(\mbox{\small $\tau_{12};\tau_1,\tau_2$})~+~ 
8 \tau_{12} \tau _3 \nonumber \\
& ~~\qquad - 4 \tau_{12} - 4 \tau_1\tau_3 - 4 \tau_2\tau_3
 \nonumber \\ 
 & ~~~~\qquad\qquad + 2 \tau_1 +2 \tau_2+ 2 \tau_3~. 
\end{align}
It is easy to verify that provided $\Lambda>2$ the original 4 degenerate solutions hold in the new combined Hamiltonian as required. 

\begin{figure}[h]{
 \includegraphics[clip,width=\columnwidth]{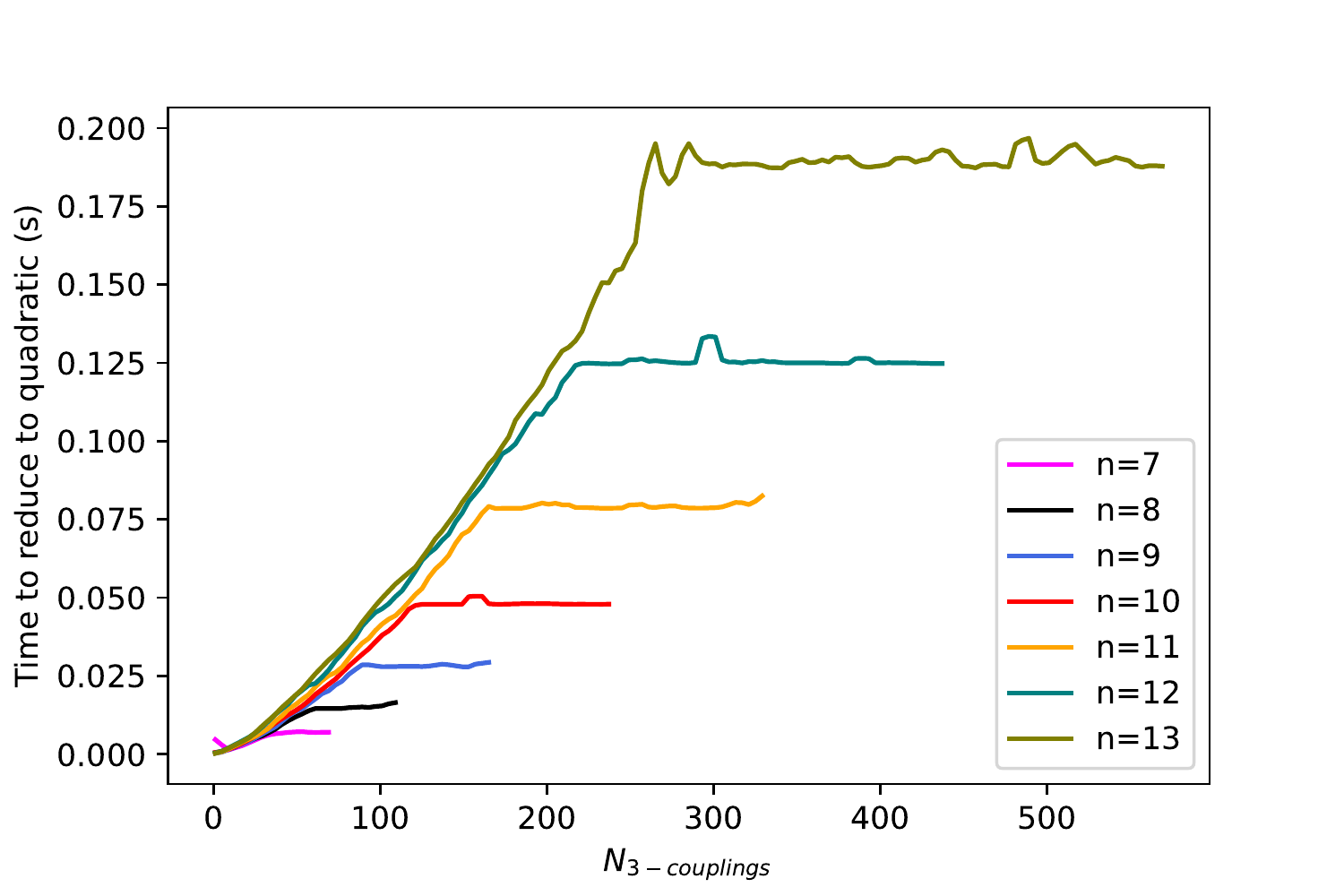}\\
 \vspace{-0.3cm} \mbox{  \bf (c)\hspace{7cm}}
}
\caption{In (a) we plot the average number of auxilliary spins required to quadratise the
Hamiltonian, versus the square root of the number of cubic interactions for different numbers of
total spins $n$. These curves exhibit a linear behavior in the central region where the slope is
given by the square root of the total number of spins, as we can see in (b). In (c) we plot the
time required to quadratise the Hamiltonian as a function of the number of cubic couplings. We
clearly see the time increasing linearly with the size of the problem.\label{fig:redplot}}
\end{figure}

With the increasing complexity of the raw Hamiltonian $\tilde{H}$ and a limited number of physical qubits at our disposal, we of course aim to find a reduction procedure that minimises the number of auxiliary variables. Therefore, the central question is how can we choose the smallest set of spin pairs that correctly collapses all the higher order terms to quadratics? In the case of cubic to quadratic, finding a spin-optimised procedure is equivalent to the set cover problem which can in turn be cast as 0-1 ILP \cite{babbush2013resource}. Both set cover and 0-1 ILP are well known to be NP-complete by analogy with vertex cover \cite{boros2002pseudo}. Therefore, generalising this to arbitrary order Hamiltonians would recast our spin-optimised problem into a task which is at least equivalent to solving $k-2$ NP-complete problems, where $k$ is the order of the Hamiltonian and therefore $k-2$ are the required layers of reduction. For this reason, we shall use a different approach based on a simple greedy algorithm which works as follows: at each reduction stage it finds the pair of binary variables $\tau_i \tau_j$ that appears most often in the Hamiltonian; wherever the pair appears, we replace $\tau_i \tau_j$ with the auxilliary logical spin $\tau_{ij}$, and add the penalty term in Eq.~\eqref{eq:constraint-Hamiltonian}. The quadratised Hamiltonian is constructed by repeating these three steps iteratively. In the language of set covering, this is equivalent to the greedy heuristic algorithm first proposed in Ref.~\cite{chvatal1979greedy}. In Fig.~\ref{fig:redplot} we have collected three plots which show how the average number of required auxilliary variables grows as we increase the number of cubic interactions, the rate of this growth in the linear central region and the time required to perform the reduction as a function of the number of cubic couplings.

\begin{figure}[h]{
  \includegraphics[clip,width=\columnwidth]{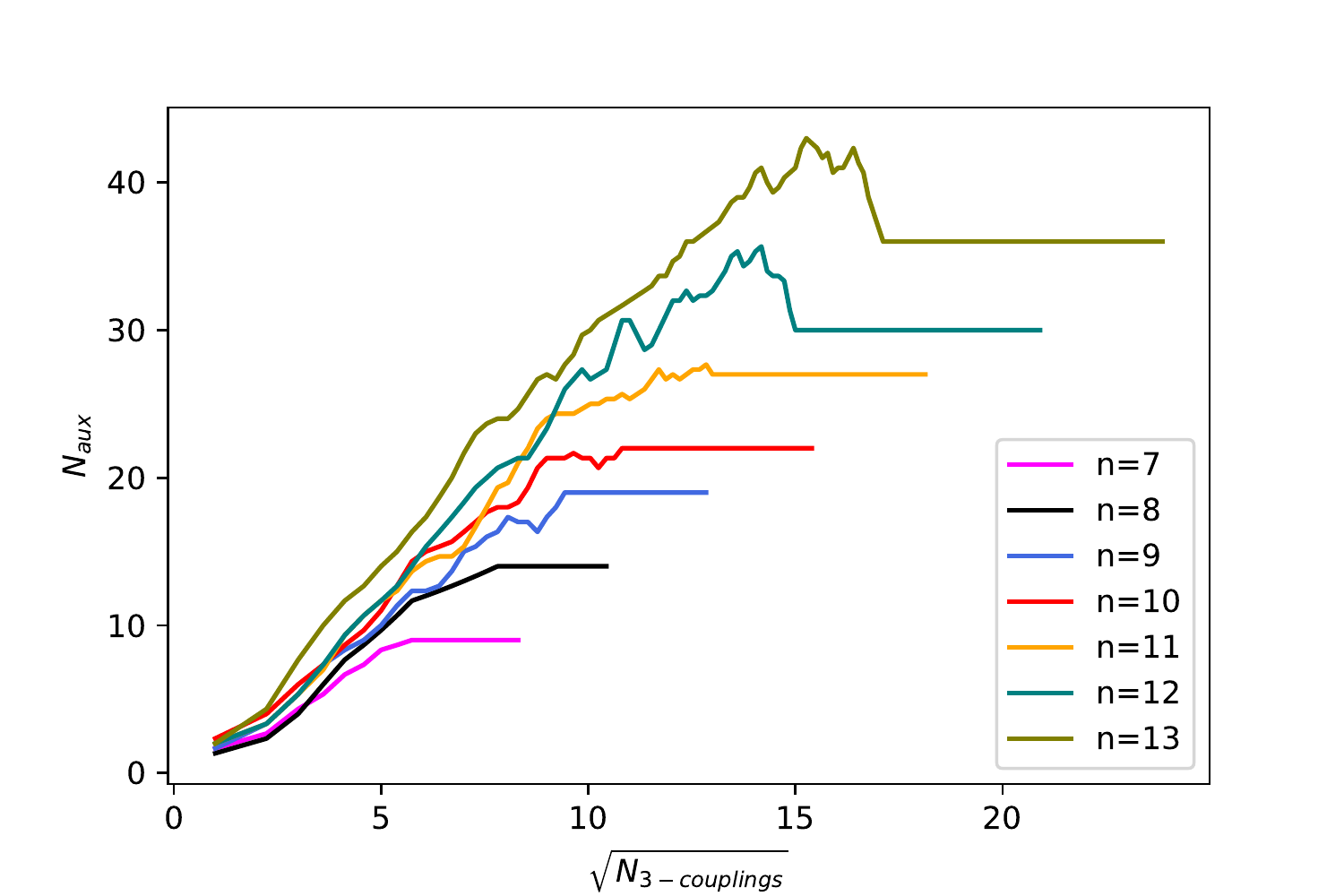}\\
 \vspace{-0.3cm} \mbox{  \bf (a)\hspace{7cm}}  }
 \end{figure}
 \begin{figure}[h]{
 \includegraphics[clip,width=\columnwidth]{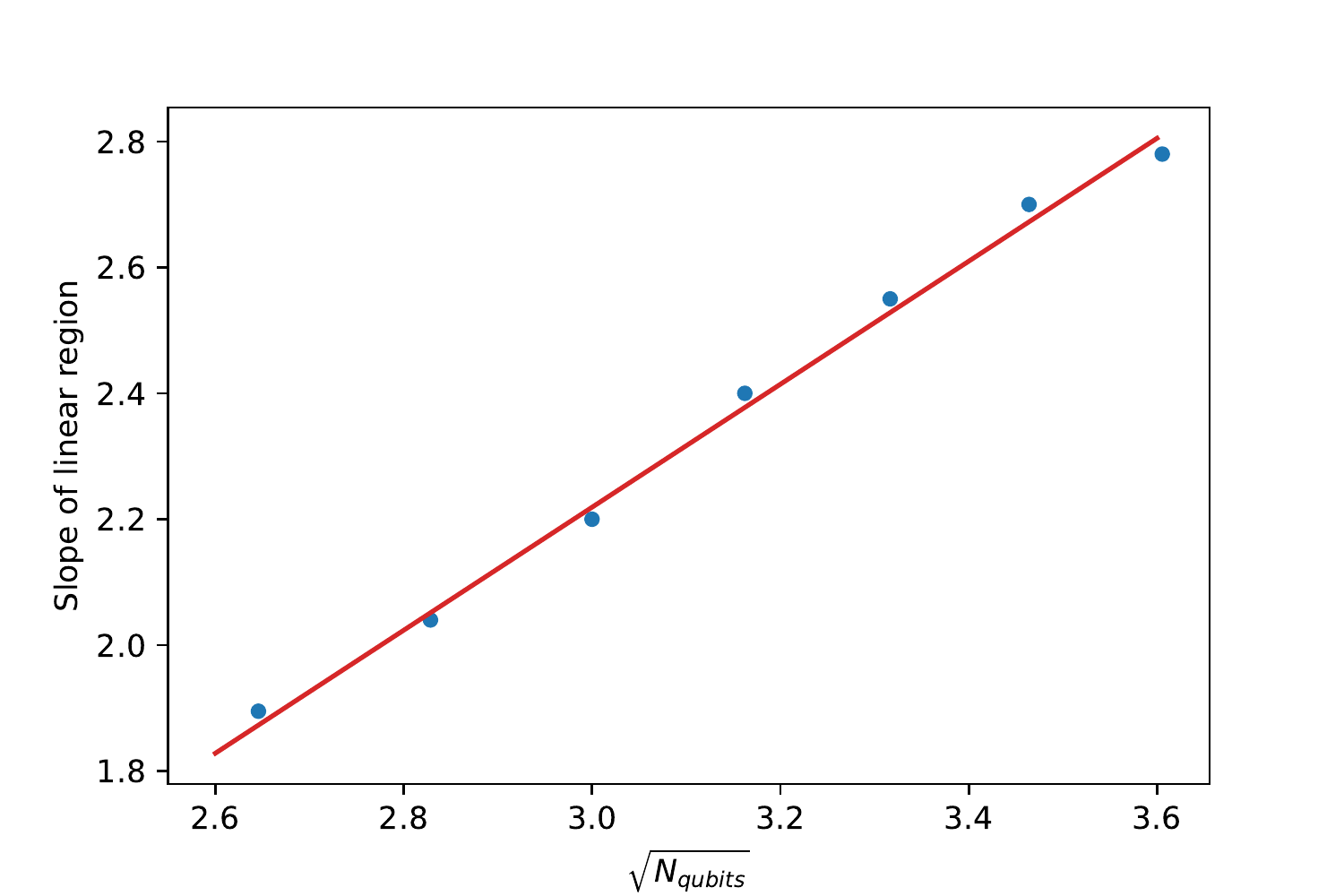}\\
 \vspace{-0.3cm} \mbox{  \bf (b)\hspace{7cm}} }
\end{figure}

These results are very similar to those obtained in Ref.~\cite{babbush2013resource} where the optimal reduction is found by solving exactly the equivalent 0-1 ILP. Two important remarks are in order. First, as we can see in Fig.~\ref{fig:redplot}a, both methods saturate approximately when the Hamiltonian contains all possible cubic interactions with $n$ qubits, namely when $N_{\text{3-couplins}} \approx \binom{n}{3}$. Second, we see that in either case the number of auxilliary spins increases linearly with the square root of the number of cubic interactions and the growth rate is given by the square root of the total number of spins, as shown in Fig.~\ref{fig:redplot}b. Nevertheless, as we can see in Fig~\ref{fig:redplot}a, in some regions our procedure uses a number of auxiliaries that is larger than the value at saturation, especially in the $n ~=~ 12$ and $n~=~13$ cases. This is of course due to the fact that we are not seeking an exact solution of the spin-optimised reduction problem. Indeed, a local optimal choice of our reduction algorithm does not necessarily lead to a global minimum in terms of the number of auxilliary spins. More precisely, in the language of the equivalent set covering problem, it has been shown in Ref.~\cite{chvatal1979greedy} that this greedy algorithm returns to an approximate solution which cannot be bigger than $H(n)$ times the minimum one, where $H(n)$ is the $n$-th harmonic number and $n$ the size of the set to be covered (namely in our case the set of all higher order couplings). However, in the problems treated below, the greedy algorithm we adopt returns quadratised Hamiltonians with at most $\sim 300$ logical spins, far below the limit imposed by for example the number of available qubits in the currently accessible quantum annealers, making it unnecessary to solve the problem exactly. Finally, we should remark that this procedure is straightforwardly generalisable to Hamiltonians of arbitrary order, requiring a number of steps which grows roughly linearly with the size of the problem, as we can see in Fig.~\ref{fig:redplot}c and also in Ref.~\cite{chvatal1979greedy}. As expected, this is in contrast with the exact method discussed in Ref.~\cite{babbush2013resource} which shows an exponentially increasing amount of time with the increasing complexity of the Hamiltonian.

\begin{figure}[t]
\includegraphics[keepaspectratio, width=0.5\textwidth]{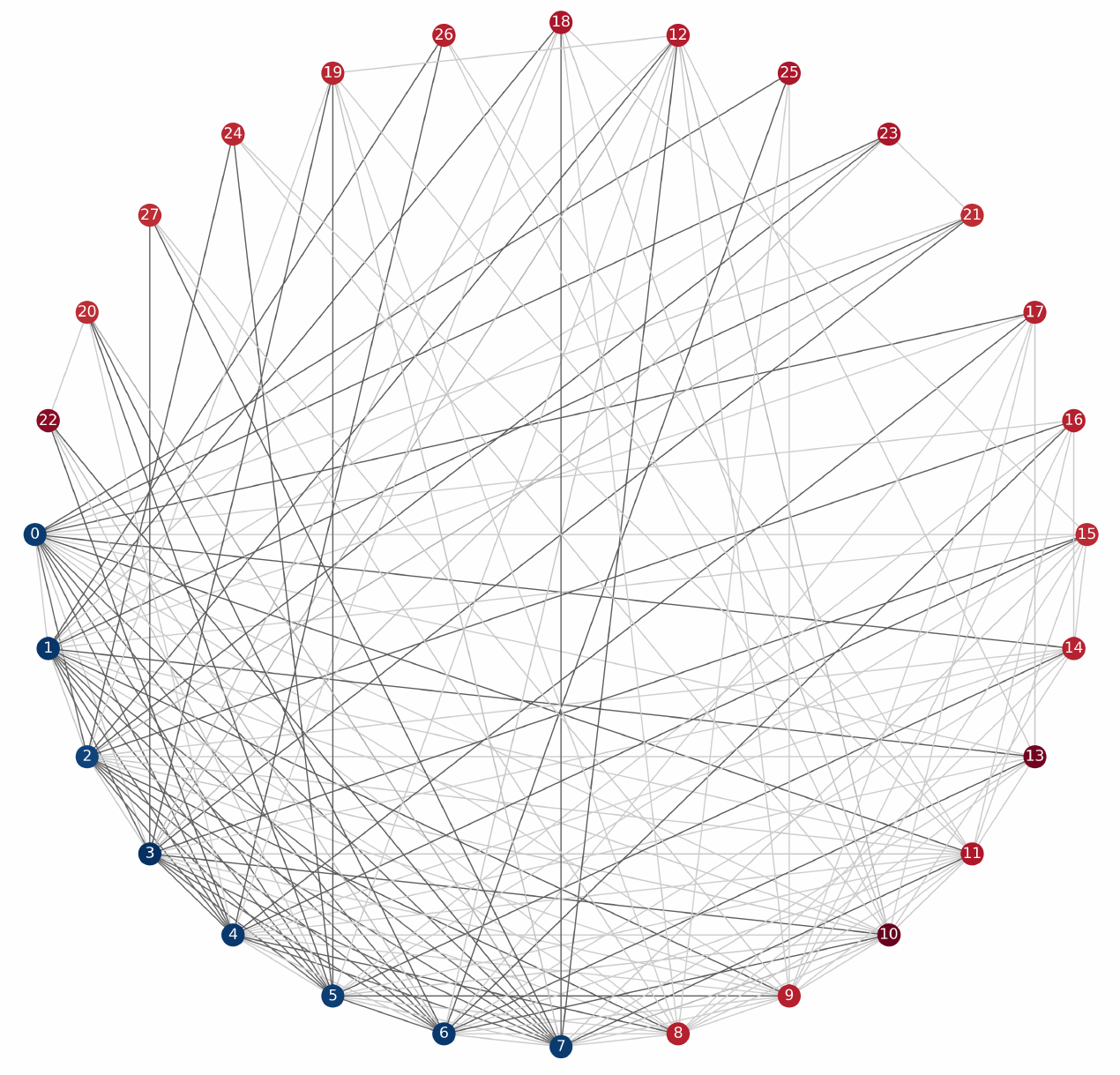}
\caption{Representation of $H ~=~ (x_1^2 + x_2^2 - 149)^2$ as a quadratised Ising model. Nodes, corresponding to spins, can be arranged in a circle, and they are quadratically coupled by the links. (Diagrams for non-quadratised models would contain junctions in the couplings).  Weaker couplings are represented in light grey, gradually getting darker for higher coupling strengths. Linear couplings are coloured from dark blue (large negative terms) to dark red (large positive terms)}
.\label{fig:reduced_graphA}
\end{figure}

Reduced Hamiltonians can be represented using connected graphs in which nodes correspond to spins and links to couplings. As an example, Fig.~\ref{fig:reduced_graphA} is a representation of the quadratised Hamiltonian associated to the first Diophantine equation in  Table~\ref{table:sol2}.

\subsection{Solution mining for improved performance}

\label{subsec:solutionmining}

Given the increase in difficulty with bitnumber, $\beta$, we will utilise a method for improving the performance. This method allows one to explore larger regions of parameter space (i.e. larger integers) without increasing $\beta$,  yielding in turn solutions with larger values. 

The method operates iteratively, by at each run constructing a brand new Hamiltonian from the previously found solutions. At say the $k$-th iteration, we minimise the Hamiltonian looking for solutions of the form
\beeq
t_i^{k} ~=~ \tau_{i,0}^{(k)} + 2\tau_{i,1}^{(k)} + s_i^k ~,
\eeeq
where $k ~=~ 0,...,N$ (with $N$ being the total number of anneal runs), and where $s^k_i$ is a classical shift that centres the new search, which is determined from a solution found in the $(k-1)$-th run: if we designate the previous solution $\hat{t}_i^{k-1}$, then the $\{s_i^k\}$ are chosen such that 
\begin{align}
&t_i^{k} ~\in~ [\hat{t}_i^{k-1}-1\,,\,\hat{t}_i^{k-1}+2] \quad \text{or} \nonumber \\ \nonumber\\
&t_i^{k} ~\in~ [\hat{t}_i^{k-1}-2\,,\,\hat{t}_i^{k-1}+1]~,
\end{align}
based on a random choice. 

This procedure finds new solutions by  
performing a kind of ``random tunnelling'' from previously found solutions (hence the name  ``solution-mining''). It generally operates well when there are many variables in the system and many different equations, because in such systems the solutions can be relatively close in each dimension of the search space (even though the total Hamming distance could be very large due to the large number of dimensions). For the two specific example problems we are discussing here, it is not a useful enhancement for finding Taxicab numbers because there one is seeking the smallest numbers, and (as we shall see) the solutions to the Diophantine system are very widely spaced. However for solving anomaly equations the method is a significant improvement. In such systems,  new solutions to the anomaly equations tend to appear with consistent frequency when the allowed charge size is increased, and it is the sheer number of anomaly equations and charges that makes the problem difficult.

It should be noted that there is no additional cost for solution-mining because even though a brand new Hamiltonian must be constructed at each stage, the embedding graph remains the same if the values of $\beta$ do not change. 
This means we construct an entirely new Hamiltonian $\tilde{H}$, but do not need to perform a new reduction of the solution. On a quantum annealer we perform reverse annealing (to be explained below) in order to collect the solution and construct the new Hamiltonian at each stage, which then simply has to be translated into new couplings via the updated $\{ s_i^k \}$ values. 

\section{\label{Sec:quantumandclassical}Quantum versus classical annealers for Diophantine equations}

In this section we shall compare quantum and classical annealers to solve Diophantine equations, using the encoding methods described above on a simple warm-up problem. It has been already shown in Ref.~\cite{chang2019quantum} that a quantum annealer can successfully be used to solve second order systems of polynomial equations. On the other hand, simulated annealing and variations thereof have been extensively applied for similar purposes (see Ref~\cite{Aguiar2014DiophantineEA,moradi2019solving}) along with other techniques such as genetic algorithms, particle swarm optimisation and so on (see for example \cite{abraham2001diophantine,abraham2010particle,abraham2013finding,perez2013numerical}). In the following we shall see that quantum annealing is also successful in solving equations of order higher than two, retrieving most of the results found using Particle Swarm Optimisation in Ref.~\cite{abraham2010particle} and Fuzzy Adaptive Simulated Annealing in Ref.~\cite{Aguiar2014DiophantineEA}. We shall find that simulated annealing on the quadratised Hamiltonian can be used to solve problems which are still hard to embed in a quantum annealer due to technological limitations (restricted number of qubits and limited connectivity). Such problems will realistically be solvable in the near future with annealers that have higher connectivity but still with at most quadratic interactions.

Let us first outline the central features of the quantum annealer that we will need for this study (for a comprehensive review see Ref.~\cite{Hauke_2020}). The quantum annealer is defined by a Hamiltonian of physical qubits of the form
\begin{align}
  \mathcal{H}(s) &~=~ A(s) \sum_\ell \sigma_{\ell,x} + \, B(s) H(\sigma_{\ell,z})  \, ,  
\end{align}
where $H$ is the problem-Hamiltonian in Eq.~\ref{eq:isingH}, $\sigma_{\ell,x}$ and $\sigma_{\ell,z}$ are the Pauli matrices acting on the $\ell^{\rm th}$ qubit, and $A(s)$, $B(s)$ are smooth functions such that $A(1) ~=~ B(0) ~=~ 0$ and $A(0) ~=~ B(1) ~=~ 1$. Meanwhile the internal couplings $h_\ell$ and $J_{\ell m}$ are fixed in each anneal session. 

Since $\mathcal{H}(1) ~=~ H(\sigma_\ell)$, the usual quantum annealing strategy is adiabatically to adjust the pre-factors $A(s)$ and $B(s)$ using the parameter $s$, such that the system ends up in a global minimum of $H$. The time dependence of $s(t)$ is defined  by the user in a so-called {\it anneal-schedule}. Thus during a {\it reverse} anneal for example one begins at $s ~=~ 1$ and with the system set in any eigenstate of $\bigotimes_\ell \sigma_{\ell, z}$. Then one allows the system to evolve quantum-mechanically by bringing $s$ to small values using a piecewise-linear function $s(t)$ of time $t$, which completes back at $s(t_{\rm final}) ~=~ 1$ when the measurement of final spins is made. 

A technological limitation is that 
the connectivity of the quantum annealing device in terms of the allowed non-vanishing couplings $J_{\ell m}$ between qubits is limited.
 Let us be specific to the architecture we will be using in this work, namely D-Wave's~\cite{LantingAQC2017} \texttt{Advantage\_system4.1}: this annealer contains 5627 qubits, connected in a \emph{Pegasus} structure, but only has a total of 40279 couplings between them.
 
The warm-up problem that we will use to compare this kind of annealer with classical simulated annealing is simply to find solutions of a generic Diophantine equation
\beeq
f(x_1,...,x_N) ~=~ 0 \,,
\eeeq
where $x_1,...,x_N \in \mathbb{Z}$ and $f: \mathbb{Z}^N \rightarrow \mathbb{Z}$ is a generic order $k$ polynomial function.

In order to solve such an equation, we again square it to form the problem-Hamiltonian:
\beeq
H(x_1,\dots,x_N) ~\equiv~ [f(x_1,\dots,x_N)]^2\,.
\eeeq
The next step is to binary encode $x_1,\dots,x_N$ as in Eq.~\eqref{eq:binary_encoding}, choosing the values of $s_i$ and $\beta$ depending on the specific problem.

The above Hamiltonian is therefore an order-$2k$ polynomial in $\tau_{i,j}$. Nevertheless, regardless of how high the order is, it can always be reduced to a quadratic  Ising Hamiltonian making use of the procedure described in Sec.~\ref{subsec:reduction}. 
Using the D-Wave's quantum annealer we find all the solutions listed in Table 2 of Ref.~\cite{abraham2010particle} (which have also been found using Fuzzy Adaptive Simulated Annealing in Ref.~\cite{Aguiar2014DiophantineEA}). We report some of them in Table~\ref{table:dwave_sol}.
\begin{table}[H]
\begin{center}
\resizebox{7.5cm}{!}{
\begin{tblr}{| c | c | c | c |  }
 \hline
 \textbf{Equation} & \textbf{Solution} & \textbf{$N_{\text{aux}}$} & \textbf{$\beta$} \\
 \hline
 $x_1^2 + x_2^2 ~=~ 625$ & 15,20 & 35 & 5 \\
 \hline
 $x_1^3 + x_2^3 ~=~ 1008$ & 2,10 & 33 & 4 \\
 \hline
 $x_1^4 + x_2^4 ~=~ 1921$ & 6,5 & 12 & 3 \\
 \hline
 \vdots &  &  \\
 \hline
 $x_1^7 + x_2^7 ~=~ 4799353$ & 9,4 & 26 & 4 \\
 \hline
 \vdots &  &  \\
 \hline
 $x_1^{15} + x_2^{15} ~=~ 1088090731$ & 4,3 & 12 & 3 \\
 \hline
\end{tblr}
}
\end{center}
\caption{A selection of Diophantine equations and corresponding solutions found using the D-Wave's quantum annealer. $N_{\text{aux}}$ is the number of auxilliary qubits necessary to quadratise the Hamiltonian, and $\beta$ is the number of qubits used to encode each variable (see Eq.~\eqref{eq:binary_encoding}).}
\label{table:dwave_sol}
\end{table}

To test this method further, we move on to Diophantine equations with increasingly many variables. As one would expect, the higher the  number of variables, the higher will be the average number of interactions per qubit. This often means that when the connectivity of the problem exceeds the native connections supported by the D-Wave Quantum Processor Unit (QPU), a single binary variable in the quadratic optimization problem needs to be represented by two (or more) qubits (called a ‘chain’) instead of one. This procedure, known as embedding is carried out by an embedding algorithm, and should be carefully monitored as it can lead to so-called {\it broken-chains} that have two or more physical qubits in the same chain taking different values. This ultimately limits the size of problems that can be solved on quantum annealers, while performing classical annealing on the same Ising Hamiltonian turns out to be successful in all the examples treated in Table~\ref{table:sol2}, where we list the equations solved using both classical (cyan) and quantum (blue) annealing.

\begin{table}[H]
\resizebox{8.7cm}{!}{
%\begin{center}
{\renewcommand{\arraystretch}{1.4}
\begin{tabular}{| c | c | c | c |}
 \hline
 \rowcolor{TopRow}
 \textbf{Equation} & \textbf{$N_{\text{aux}}$} & \textbf{$\beta$} & \textbf{Solutions} \\
 \hline
 \rowcolor{lblue}
 $x_1^2 + x_2^2 ~=~ 149$  & 20 & 4 & (7, 10)\\
 \hline
 \rowcolor{lblue}
 $x_1^2 + x_2^2 + x_3^3 ~=~ 244$  & 18 & 4 & (12, 8, 6) \\
 \hline
 \rowcolor{lblue}
 $x_1^2 + \dots + x_4^2 ~=~ 295$  & 24 & 4 & (14, 7, 1, 7), (11, 5, 7, 10)\\
 \hline
 \rowcolor{lblue}
 $x_1^2 + \dots + x_5^2 ~=~ 325$  & 30 & 4 & (4, 8, 8, 9, 10)\\
 \hline
 \rowcolor{lblue}
 $x_1^2 + \dots + x_6^2 ~=~ 420$  & 36 & 4 & (1, 11, 9, 10, 6, 9)\\
 \hline
 \rowcolor{cyan}
 $x_1^2 + \dots + x_7^2 ~=~ 450$ & 70 & 5 & \begin{tabular}{@{}c@{}}(10, 4, 17, 2, 6, 1, 2) \\ (2, 7, 2, 2, 18, 8, 1) \\ \dots \end{tabular} \\
 \hline
 \rowcolor{cyan}
 $x_1^2 + \dots + x_8^2 ~=~ 590$  & 80 & 5 & \begin{tabular}{@{}c@{}} (9, 11, 1, 6, 5, 14, 3, 11)\\ (1, 11, 8, 9, 4, 15, 1, 9) \\ \dots \end{tabular} \\
 \hline
 \rowcolor{cyan}
 $x_1^2 + \dots + x_9^2 ~=~ 720$  & 90 & 5 & \begin{tabular}{@{}c@{}}(14, 2, 13, 8, 9, 6, 8, 9, 5)\\ (7, 16, 1, 1, 20, 2, 1, 2, 2) \\ \dots \end{tabular}\\
 \hline
 \rowcolor{cyan}
 $x_1^2 + \dots + x_{10}^2 ~=~ 956$  & 100 & 5 & \begin{tabular}{@{}c@{}}(12, 2, 8, 20, 3, 11, 11, 8, 2, 5)\\ (3, 5, 7, 5, 6, 4, 2, 14, 14, 20) \\ \dots \end{tabular}  \\
 \hline
 \rowcolor{cyan}
 $x_1^2 + \dots + x_{11}^2 ~=~ 1502$  & 110 & 5 & \begin{tabular}{@{}c@{}}(6, 2, 9, 22, 9, 5, 12, 23, 1, 6, 9)\\ (9, 7, 9, 23, 1, 5, 1, 23, 11, 9, 2) \\ \dots \end{tabular} \\
 \hline
 \rowcolor{cyan}
 $x_1^2 + \dots + x_{12}^2 ~=~ 3842$ & 120 & 5 & \begin{tabular}{@{}c@{}}(1, 21, 4, 6, 12, 2, 28, 17, 18, 29, 31, 1)\\ (17, 8, 17, 32, 28, 5, 19, 16, 17, 14, 16, 3)
 \\ \dots \end{tabular} \\
 \hline
\end{tabular}
}
%\end{center}
}
\caption{A list of Diophantine equations solved using both quantum (blue) and classical (cyan) annealing.}
\label{table:sol2}
\end{table}
This reproduces the results in Refs.~\cite{abraham2010particle,Aguiar2014DiophantineEA} obtained using Particle Swarm Optimisation and Fuzzy Adaptive Simulated Annealing respectively.

\section{\label{Sec:TaxiCab}Ramanujan (TaxiCab) numbers}

Having demonstrated that problems such as these can in principle be already solved on a quantum annealer, we are now ready to move on to the more complicated class of problems outlined in the introduction, beginning in this section with Taxicab numbers. In line with the above discussion, classical simulated annealing turns out to be superior currently for these problems, so we will use that annealing method in this and the next section.

In general, finding Taxicab numbers is not a trivial task and for higher Taxicabs, such as $\text{Ta}(7)$, $\text{Ta}(8)$ etc., only an upper bound is known \cite{Boyer}. Indeed (using the  $(k,m,n)$ notation for these numbers), it is interesting to note that no $(5,2,2)$ numbers have been found, despite searches up to $10^{26}$ (see Ref.~\cite{Guy:1994a}). 

Let us show explicitly how we use the reduction technique of Subsection~\ref{subsec:reduction} to construct an Ising Model Hamiltonian whose ground states are precisely the Taxicab numbers we want to find. As a first example we focus on $\text{Ta}(2)$, i.e. we want to find four non-negative integer numbers such that 
\beeq
a^3 + b^3 ~=~ c^3 + d^3\,,\quad a \neq \{c,d\}\,.
\label{eq:taxicab2}
\eeeq
We again use binary encoding (see Eq.~\eqref{eq:binary_encoding}) with $\beta ~=~ 5$, $s_i ~=~ 1$ (numbers from $1$ to $32$) and $t_i \in \{a,b,c,d\}$. To impose the equality between the two sums of cubes we define the following Hamiltonian
\beeq
H_D ~=~ (a^3 + b^3 - c^3 - d^3)^2\, . 
\label{eq:hamiltonianD}
\eeeq
However, this is not the end of the story as must also encode the constraint $a \neq \{c,d\}$  to avoid all the trivial minima of the above Hamiltonian, which occur when $a = c$ and $b=d$ or vice versa. 
In other words we want to construct the $H_C$  Hamiltonian such that it has its global minimum when $a \neq c$ and $a \neq d$. It is more straightforward to write such a constraint Hamiltonian directly in terms of binary variables $\tau_{i,k}$, where $i \in \{a,b,c,d\}$ and $k = 0,...,\beta-1$. It is easy to see that the
Hamiltonian
\begin{align}
    H_C ~&\equiv~ H_{\delta}(a,c) + H_{\delta}(a,d) \nonumber \\
    & \qquad \equiv  \prod_{k=0}^{\beta-1}\left(1-(\tau_{a,k}-\tau_{c,k})^2\right) \nonumber\\ &\qquad \qquad +\prod_{k=0}^{\beta-1}\left(1-(\tau_{a,k}-\tau_{d,k})^2\right)
\end{align}
achieves this.
Explicitly, one finds that
\beeq
H_C ~=~ \begin{cases} 0, \text{   when } a \neq c \text{  and  } a \neq d \, , \\ 
1, \text{   when } a ~=~ c \text{  and  } a \neq d \, , \\
1, \text{   when } a ~=~ d \text{  and  } a \neq c \, , \\
2, \text{   when } a ~=~ c ~=~ d \, .
\end{cases}
\eeeq
The Hamiltonian we shall use is then the sum
\beeq
\tilde{H} ~=~ H_D + H_C\, .
\eeeq
Written is terms of $\tau$'s, this Hamiltonian is a polynomial of order $2\beta$ for $\beta ~\geq~ 3$. Again, setting $\beta = 5$ and using the technique described in Sec.~\ref{Sec:Annealer} we can reduce it to a quadratic Hamiltonian by adding $98$ auxilliary spins.

\begin{figure}[H]
\centering
\includegraphics[keepaspectratio, width=0.5\textwidth]{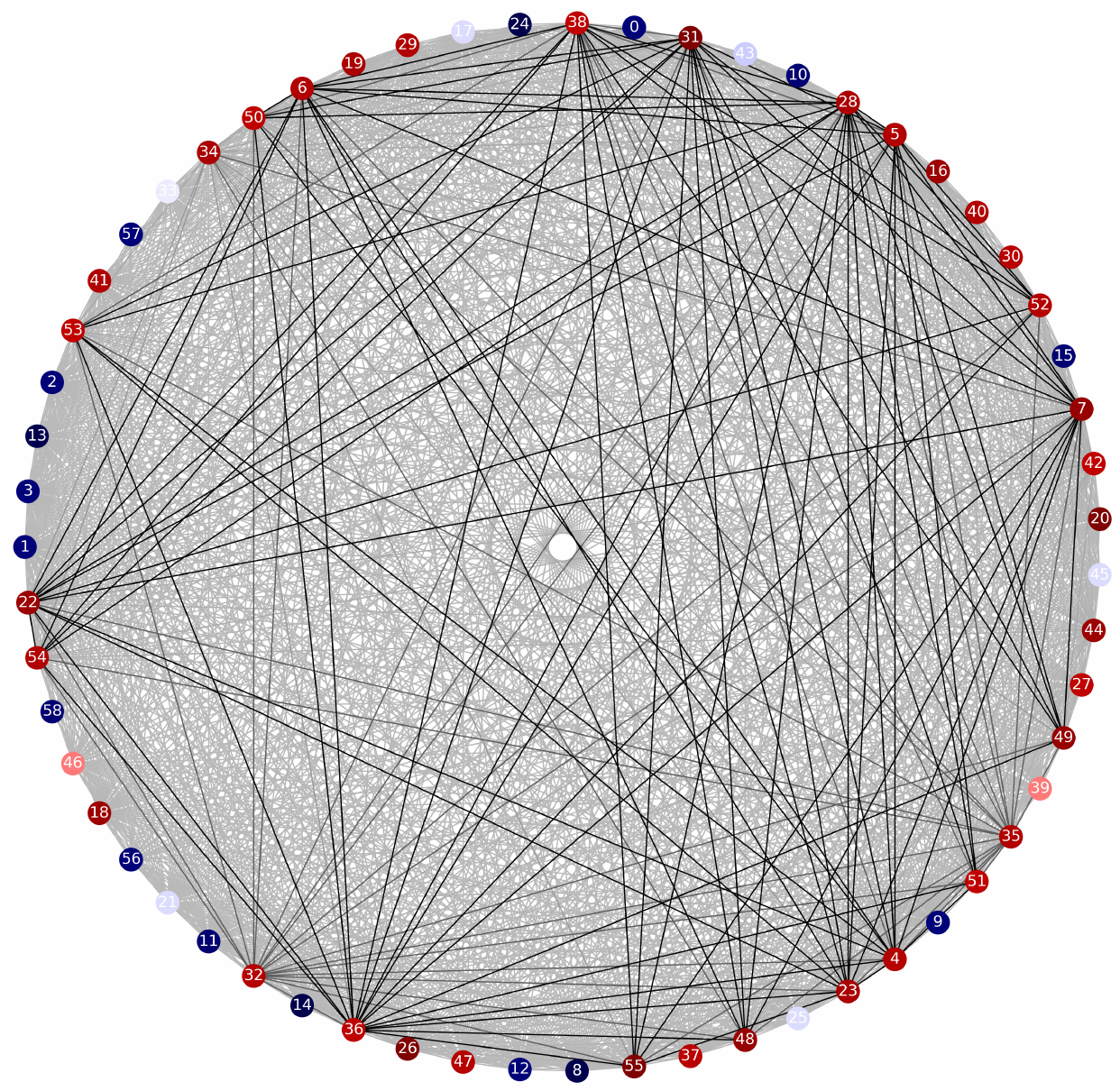}
\caption{Representation of the  Ising Hamiltonian corresponding to the Ta$(2)$ problem.}
\label{fig:reduced_graphB}
\end{figure}
In Fig.~\ref{fig:reduced_graphB} we  represent the reduced Hamiltonian for $\beta = 4$. We see that stronger couplings are rare among the interactions, which mostly form a very complex network of weaker couplings in the background. Classical annealing on the reduced Hamiltonian yields all the solutions written in Eq.~\eqref{eq:taxi_solutions}, namely all the Taxicab numbers with $a,b,c,d ~\leq~32$.

Let us now push this further and attempt to solve more complicated generalisation of the Taxicab problem, $(k,m,n)$ where
\begin{align}
(k,m,n) ~&\equiv~  a_{1}^k \, +\, ... \,+ \, a_{m}^k  \nonumber \\
& = ~ b_{1}^k \,+\, ...\, + \, b_{n}^k, 
\end{align}
where $\{a_{1},...,a_{m}\} \neq \{b_{1},...,b_{n}\}$. 
Beginning with $(4,3,3)$ numbers,
\begin{align}
(4,3,3) ~&=~  a^4 \, +\, b^4 \, + \, c^4  \nonumber \\
& = ~ d^4 \,+\, e^4 \, + \, f^4 \,, 
\label{eq:432}
\end{align}
we define the following Hamiltonians
\beeq
H_D ~=~ (a^4 + b^4 + c^4 - d^4 - e^4 - f^4)^2\, , 
\label{eq:hamiltonianD2}
\eeeq
and
\beeq
H_C ~=~ H_\delta(a,d) + H_\delta(a,e) + H_\delta(a,f)\, ,
\label{eq:hamiltonianC2}
\eeeq
to impose the equality in Eq.~\eqref{eq:432} and also to force $a \neq d,e,f$.
The order of the complete Hamiltonian, which is the sum of Eq.~\ref{eq:hamiltonianC2} and Eq.~\ref{eq:hamiltonianD2}, is $2\beta$ for $\beta \geq 4$. Again, it can be reduced to a quadratic one by adding $66$ auxilliary variables in the case with $\beta = 4$ and $154$ in the case with $\beta = 5$. Several anneal runs (each with $10000$ reads) with $\beta = 4,5$ yield the following results
\begin{table}[H]
\begin{center}
\begin{tabular}{| c | c | c | c | c | c | c | }
 \hline
 $(4,3,3)$ & $a$ & $b$ & $c$ & $d$ & $e$ & $f$  \\
 \hline
 2673 & 3 & 6 & 6 & 7 & 2 & 4  \\
  \hline
 16562 & 9 & 1 & 10 & 11 & 6 & 5 \\
  \hline
 28593 & 2 & 13 & 2 & 9 & 6 & 12 \\
  \hline
 35378 & 13 & 4 & 9 & 11 & 12 & 1 \\
  \hline
 43218 & 11 & 13 & 2 & 14 & 7 & 7 \\
  \hline
 54977 & 4 & 8 & 15 & 9 & 14 & 10 \\
  \hline
 195122 & 21 & 5 & 2 & 9 & 13 & 20 \\
  \hline
 324818 & 14 & 9 & 23 & 21 & 2 & 19 \\
  \hline
 619337 & 28 & 8 & 5 & 7 & 26 & 20 \\
  \hline
 847602 & 1 & 25 & 26 & 29 & 19 & 10 \\
  \hline
 1071713 & 12 & 32 & 7 & 28 & 26 & 3 \\
  \hline
 1178898 & 29 & 11 & 26 & 1 & 32 & 19 \\
  \hline
 1328498 & 29 & 9 & 28 & 23 & 32 & 3 \\
 \hline
\end{tabular}
\end{center}
\caption{List of $(4,3,3)$ numbers found using $\beta ~=~ 3,4,5$ and $10000$ reads per anneal run.}
\label{table:4,3,3}
\end{table}
To comment on the efficacy of the method: the search space is of order $32^6\sim 10^9$, and yet these solutions are found after order $10^5$ reads.

For the remainder of this section we consider the $(3,n,m)$ numbers, where $n,m \in \mathbb{N^+}$. For this purpose we define the following Hamiltonian
\beeq
\tilde{H} ~=~ \left(\sum_{i=1}^n a_i^3 - \sum_{i=1}^m b_i^3\right)^2 \, , 
\label{eq:Hamiltonian3NN}
\eeeq
where in this case we do not add any constraint Hamiltonian to enforce $\{a_i\} \neq \{b_i\}$ when $n=m$, because here it is sufficient to simply check at the end of each anneal run if the minimum is trivial or not.

Tables~\ref{table:3,1,5}-\ref{table:3,1,7} list some of the solutions found for $n = 1$ and $m = 5,7$ with $\beta = 5,6$. Cases with $n = m \equiv N$, with $N = 6,7,8$ are listed in Tables~\ref{table:3,6,6}-\ref{table:3,8,8}.

%$120$ and $160$ auxiliary spins are needed to reduce the Hamiltonian in the $\beta = 5$ case with %$m=5,7$ respectively. $216$ and $288$ in the $\beta = 6$ case. \\
%Here are a few $(3,1,5)$ numbers
\begin{table}[H]
\begin{center}
\begin{tabular}{| c | c | c | c | c | c | c | }
\hline
(3,1,5) & $a_1$ & $b_1$ & $b_2$ & $b_3$ & $b_4$ & $b_5$ \\
 \hline
729 & 9 & 1  & 3 &  4 & 5 & 8   \\
 \hline
1728 & 12 & 3 & 10 & 4 & 8 & 5  \\
 \hline
68921 & 41 & 3 & 17 & 21 & 28 & 32   \\
 \hline
125000 & 50 & 2 & 8 & 24 & 36 & 40  \\
 \hline
185193 & 57 & 16 & 17 & 30 & 40 & 44  \\
 \hline
216000 & 60 & 11 & 16 & 25 & 45 & 47  \\
 \hline
262144 & 64 & 9 & 18 & 31 & 44 & 52 \\
 \hline
\end{tabular}
\end{center}
~\caption{A list of $(3,1,5)$ numbers found using $\beta ~=~ 5,6$. The reduction needs $120$ and $216$ auxilliary spins respectively.\label{table:3,1,5}}
\end{table}

\begin{table}[H]
\begin{center}
\begin{tabular}{| c | c | c | c | c | c | c | c | c | }
\hline
(3,1,7) & $a_1$ & $b_1$ & $b_2$ & $b_3$ & $b_4$ & $b_5$ & $b_6$ & $b_7$ \\
\hline
2744  & 14 & 2 & 3 & 5 & 7 & 8 & 9 & 10 \\
\hline
13824 & 24 & 3 & 5 & 8 & 9 & 13 & 15 & 19  \\
\hline
32768 & 32 & 1 & 6 & 15 & 16 & 17 & 20 & 23  \\
\hline
148877 & 53 & 3 & 21 & 24 & 28 & 29 & 32 & 36   \\
\hline
205379 & 59 & 5 & 12 & 13 & 18 & 23 & 43 & 47  \\
\hline
238328 & 62 & 17 & 20 & 22 & 31 & 32 & 38 & 46 \\
\hline
\end{tabular}
\end{center}
\caption{A list of $(3,1,7)$ numbers found using $\beta = 5,6$. The reduction needs $160$ and $288$ auxilliary spins respectively.\label{table:3,1,7}}
\end{table}
\begin{table}[H]
\begin{center}
\begin{tabular}{| c | c | c | c | c | c | c | c | c | c | c | c | c |}
 \hline
$(3,6,6)$ & $a_1$ & $a_2$ & $a_3$ & $a_4$ & $a_5$ & $a_6$ & $b_1$ & $b_2$ & $b_3$ & $b_4$ & $b_5$ & $b_6$  \\
 \hline
5012 & 2 & 4 & 5 & 7 & 12 & 14 & 3 & 6 & 8 & 9 & 11 & 13  \\
 \hline
7975 & 1 & 7 & 8 & 10 & 14 & 15 & 3 & 4 & 9 & 11 & 12 & 16 \\
 \hline
8309 & 1 & 5 & 7 & 10 & 14 & 16 & 4 & 6 & 9 & 12 & 13 & 15 \\
 \hline
41873 & 3 & 8 & 11 & 14 & 26 & 27 & 1 & 6 & 13 & 19 & 22 & 28 \\
 % \hline
%44207 & 5 &  6 &  8 &  9 &  25 &  30 &  15 &  16 &  17 &  18 &  23 &  24 \\
 \hline
48438 & 9 &  13 &  17 &  20 &  22 &  28 & 1 &  4 &  15 &  18 &  23 &  30 \\
\hline
51318 & 1 &  10 &  15 &  17 &  21 &  32 & 3 &  5 &  9 &  16 &  28 &  29 \\
 \hline
52359 & 5 &  6 &  7 &  11 &  26 &  32 & 3 &  14 &  15 &  20 &  24 &  29 \\
% \hline
%71407 &  13 &  16 &  17 &  25 &  26 &  30 & 1 &  3 &  15 &  24 &  29 &  31 \\
 \hline
78730 & 2 & 3 & 23 & 26 & 28 & 30 & 11 & 14 & 16 & 20 & 31 & 32 \\
\hline
86400 & 3 & 9 & 21 & 24 & 31 & 32 & 4 & 5 & 26 & 27 & 28 & 30 \\
\hline
\end{tabular}
\end{center}
\caption{$(3,6,6)$ solutions with $\beta = 4,5$. The reduction needs $120$ and $240$ auxilliary spins respectively.\label{table:3,6,6}}
\end{table}

\begin{table}[H]
\begin{center}
\begin{tabular}{| c | c | c | c | c | c | c | c | c | c | c | c | c | c | c |}
 \hline
$(3,7,7)$ & $a_1$ & $a_2$ & $a_3$ & $a_4$ & $a_5$ & $a_6$ & $a_7$ & $b_1$ & $b_2$ & $b_3$ & $b_4$ & $b_5$ & $b_6$ & $b_7$  \\
 \hline
 39256 & 3 & 8 & 9 & 14 & 17 & 22 & 27 & 2 & 5 & 11 & 16 & 19 & 21 & 26 \\
 \hline
45063 & 3 & 5 & 7 & 13 & 14 & 19 & 32 & 2 & 9 & 10 & 15 & 18 & 23 & 28  \\
\hline
46411 & 7 & 9 & 14 & 17 & 18 & 23 & 27 & 3 & 5 & 8 & 10 & 11 & 22 & 32 \\
\hline
52094 & 1 & 6 & 13 & 17 & 22 & 23 & 28 & 3 & 7 & 14 & 16 & 18 & 21 & 31 \\
\hline
%61173 & 3 & 7 & 9 & 22 & 23 & 26 & 27 & 1 & 2 & 15 & 17 & 21 & 24 & 31 \\
%\hline
63224 & 7 & 9 & 12 & 18 & 20 & 24 & 32 & 2 & 3 & 13 & 14 & 19 & 29 & 30 \\ 
\hline
73276 & 6 & 12 & 14 & 15 & 24 & 29 & 30 & 9 & 17 & 18 & 20 & 23 & 27 & 28 \\
\hline
77687 & 2 & 9 & 17 & 21 & 24 & 28 & 30 & 4 & 5 & 15 & 16 & 26 & 27 & 32 \\
\hline
\end{tabular}
\end{center}
\label{table:3,7,7}
\caption{$(3,7,7)$ solutions with $\beta = 5$. The reduction needs $280$ auxilliary spins.}
\end{table}

\begin{table}[H]
\begin{center}
\begin{tabular}{| c | c | c | c | c | c | c | c | c | c | c | c | c | c | c | c | c |}
 \hline
$(3,8,8)$ & $a_1$ & $a_2$ & $a_3$ & $a_4$ & $a_5$ & $a_6$ & $a_7$ & $a_8$ & $b_1$ & $b_2$ & $b_3$ & $b_4$ & $b_5$ & $b_6$ & $b_7$ & $b_8$  \\
\hline
50139 & 1 & 3 & 6 & 10 & 12 & 20 & 23 & 30 & 2 & 5 & 9 & 13 & 17 & 19 & 25 & 27 \\
\hline
73206 & 1 & 3 & 4 & 17 & 20 & 25 & 26 & 30 & 5 & 8 & 9 & 10 & 19 & 21 & 28 & 32 \\
\hline
78202 & 3 & 4 & 17 & 18 & 19 & 24 &  27 & 30 & 1 & 2 & 9 & 16 & 20 & 22 & 28 & 32 \\
\hline
85418 & 2 & 3 & 9 & 16 & 18 & 23 & 31 & 32 & 6 & 10 & 14 & 15 & 24 & 26 & 27 & 30 \\ 
\hline
\end{tabular}
\end{center}
\caption{$(3,8,8)$ solutions with $\beta = 5$. The reduction needs $320$ auxilliary spins. Note that even the smallest $(3,8,8)$ number represents a solution in a search space of size $32^{16}\sim 10^{24}$.\label{table:3,8,8}}
\end{table}

Note that all the above solutions are non-trivial $(3,n,m)$ numbers, in that they are not sums of smaller solutions. Indeed, it may happen that a $(3,n,m)$ number is actually the sum of $(3,p,q)$ and $(3,k,l)$ numbers with $p+k ~=~ n$ and $q+l ~=~ m$. In order to avoid such trivial solutions, we have simply removed them by hand at the end of each anneal run.
    
\section{\label{Sec:Anomalies}Anomaly cancellation in the Standard Model with an extra $U(1)$}

Having road-tested our reduction methods on simple problems, we now turn to a physical application, namely the anomaly cancellation conditions in the Standard Model extended by an extra $U(1)$ gauge symmetry. This is one of the simplest and most studied extensions of the Standard Model (see Ref.~\cite{Langacker:2009} for a review of $Z'$ physics), and it has been the target of numerous experimental searches \cite{zyla2020}. 

The generalities of anomaly cancellation for such systems have been discussed in Refs.~\cite{ Appelquist:2003, Batra:2005rh,liu2012searching,ismail2017axial,Allanach:2018vjg,tang2018flavor,Ellis:2018, Allanach:2019uuu,Costa:2019anomaly, Smolkovivc2019anomaly,Allanach:2020geometric, Costa:2020,Allanach:2020zna,Allanach:2021bfe,Podo:2022gyj}. In this work we will for concreteness specialise to the models studied in Ref.~\cite{Allanach:2020zna}. Here, the main assumption is that the chiral fermions appear in the usual 3 families of quarks and leptons, together with 3 right-handed neutrinos. The charges under the additional $U(1)$ are labelled by $\{Q_i, U_i, D_i, L_i, E_i, N_i\}$, respectively, with $i \in \{1, 2, 3\}$ indicating the generation number. Under this assumption the anomaly cancellation condition yields the following set of Diophantine equations for the charges:
\begin{align} 
&\sum_{i=1}^3\, (6Q_i + 3U_i + 3D_i + 2L_i + E_i + N_i) ~=~ 0\, , \label{eq:allanach1} \\ 
&\sum_{i=1}^3\,(3Q_i + L_i) ~=~ 0\, , \label{eq:allanach2}\\ 
&\sum_{i=1}^3\, (2Q_i + U_i + D_i) ~=~ 0\, , \label{eq:allanach3}\\ 
&\sum_{i=1}^3\, (Q_i + 8U_i + 2D_i + 3L_i + 6E_i) ~=~ 0\, , \label{eq:allanach4}\\
&\sum_{i=1}^3\, (Q_i^2 - 2U_i^2 + D_i^2 - L_i^2 + E_i^2) ~=~ 0\, , \label{eq:allanach5}\\ 
&\sum_{i=1}^3\, (6Q_i^3 + 3U_i^3 + 3D_i^3 + 2L_i^3 + E_i^3 + N_i^3) ~=~ 0\,. 
\label{eq:allanach6}
\end{align}
A general solution to the above equations has already been found analytically in Ref.~\cite{Allanach:2020zna}. However, we shall demonstrate here that these problems can be also tackled using Ising model annealing (in practice here we use simulated annealing, but ultimately quantum annealers will be practicable). 

As for the Taxicab problem, we begin constructing the Hamiltonian by simply squaring and summing the left hand side of all the above equations. We encode all the variables involved as in Eq.~\eqref{eq:binary_encoding} with $t_i \in \{Q_i, U_i, D_i, L_i, E_i, N_i\}$ and take $s_i = -1$ for all the charges. We set $\beta = 2$, thus looking for solutions with entries from $-1$ to $2$. Note that although the number of bits we use to represent each variable is relatively low, the number of possible configurations of these $3 \times 6 ~=~ 18$ charges with possible values in $[-1,2]$ is already quite high: $4^{18} \sim 10^{10}$. It is worth mentioning that in this particular case a comprehensive scan can be completed with far fewer attempts due to generation permutation symmetry in the equations. Indeed, it is easy to see that the anomaly equations are invariant under arbitrary permutations of $\{A_1,A_2,A_3\}$, where $A \in \{Q,U,D,L,E,N\}$, giving an $(S_3)^6$ permutational symmetry that could be exploited if we were looking for solutions by exhaustive scanning over all the different configurations.

Of course our goal here is to avoid using such tricks, but to instead find solutions using annealing on the reduced Ising Hamiltonian. For $\beta = 2$ the reduction requires only $18$ auxiliaries. We have performed several anneal runs with $10000$ reads obtaining an average of $60$ distinct solutions per anneal run. In the following table we present a sample of three of them.
\begin{table}[H]
\resizebox{8.5cm}{!}{
\begin{tabular}{| c | c | c | c | c | c | c | c |c | c | c | c | c | c |c | c | c | c |}
 \hline
 $Q_1$ & $Q_2$ & $Q_3$ & $U_1$ & $U_2$ & $U_3$ & $D_1$ & $D_2$ & $D_3$ & $L_1$ & $L_2$ & $L_3$ & $E_1$ & $E_2$ & $E_3$ & $N_1$ & $N_2$ & $N_3$ \\
 \hline
 1 & 0 & -1 & -1 & 1 & 0 & 1 & -1 & 0 & -1 & 0 & 1 & -1 & 0 & 1 & 0 & 1 & -1  \\
 \hline
 -1 &  1 &  0 &  0 &  -1 &  1 &  -1 &  1 &  0 &  -1 &  0 &  1 &  0 &  1 &  -1 &  1 &  0 &  -1  \\
 \hline
 1  & -1  & 0  & -1  & 1  & 0  & 1  & 0  & -1  & 1  & 0  & -1  & -1  & 1  & 0  & -1  & 0  & 1    \\
 \hline
\end{tabular}
} 
\caption{A sample of three solutions found using $\beta ~=~ 2$ and $10000$ reads in each anneal run.}
\label{table:allanach_1}
\end{table}
One might expect higher values of $\beta$ to lead to new solutions with bigger entries, along with those previously found. However, for this specific problem, classical annealing turns out to be unfruitful for $\beta > 2$. To explain why it is useful to inspect how the energy gap $\Delta$ between the ground state and the first excited state scales as a function of the size of the problem. It can be shown (see Ref.~\cite{chang2019least}) that
\beeq
\Delta ~\sim~ \mathcal{O} \left( \frac{2^{-n \mu}}{m \alpha} \right)~,
\eeeq
where $\alpha$ and $\mu$ are the number of additions and multiplications respectively in the Hamiltonian written in terms of binary variables, $m$ is the number of equations we want to solve and $n$ is the the effective precision, which is the difference between the largest and smallest nonzero absolute values representable among all the variables in the system. Increasing $\beta$ makes all these parameter bigger, including $n$ and $\mu$, causing an exponential shrinkage of the energy gap between the ground states and the first excited states, which in turn considerably affects the algorithm's performance.

To improve our results and find solutions with bigger entries we use the solution-mining method described in Sec.~\ref{subsec:solutionmining}. After $30$ anneal runs this yields $153$ solutions with entries between $-13$ and $13$. Note that a complete scan on all possible such configurations, even exploiting the $(S_3)^6$ permutational symmetry, would involve $\binom{13 \times 2 + 1}{3}^6 \sim 10^{20}$ trials, which is  infeasible with conventional computing methods. It should be noted that we do not make use of the permutational symmetry and the Ising machine is in principle succeeding within a  search space of $26^{18} \sim 3\times 10^{25}$, although it is not yet clear how exhaustive the method of small $\beta$ plus solution mining can eventually be.

In Table~\ref{table:allanach_2} we present a sample of ten of the solutions found.
\begin{table}[H]
\resizebox{8.5cm}{!}{
\begin{tabular}{| c | c | c | c | c | c | c | c |c | c | c | c | c | c |c | c | c | c |}
 \hline
 $Q_1$ & $Q_2$ & $Q_3$ & $U_1$ & $U_2$ & $U_3$ & $D_1$ & $D_2$ & $D_3$ & $L_1$ & $L_2$ & $L_3$ & $E_1$ & $E_2$ & $E_3$ & $N_1$ & $N_2$ & $N_3$ \\
 \hline
 -1 & 0 & 1 & -1 & 0 & 1 & 1 & -1 & 0 & 1 & 0 & -1 & 0 & -1 & 1 & 1 & 0 & -1 \\
\hline
 0 & -2 & 2 & 1 & -1 & 2 & -2 & 0 & 0 & 0 & 1 & -1 & 0 & -1 & -1 & 0 & 1 & 1   \\
 \hline
 3  & -1  & -2  & -1  & -2  & 3  & -4  & 2  & 2  & 0  & -3  & 3  & 2  & -2  & 0  & 2  & -3  & 1   \\
 \hline
 3  & -2  & -1  & 1  & -3  & 3  & -4  & 3  & 0  & -1  & 0  & 1  & -1  & 0  & 0  & 3  & -3  & 1 \\
\hline
-1  & 1  & 0  & -2  & -1  & 4  & -5  & 4  & 0  & -2  & -1  & 3  & 0  & 2  & -3  & 1  & -2  & 2 \\
\hline
1  & -1  & 0  & 0  & -2  & 5  & -6  & 4  & -1  & -1  & 0  & 1  & 0  & -1  & -2  & 0  & -2  & 5 \\
\hline
1  & 0  & 0  & -1  & -2  & 6  & -7  & 4  & -2  & -3  & 0  & 0  & 2  & 1  & -4  & 0  & 0  & 7 \\
\hline
2  & -1  & -1  & 2  & -3  & 4  & -6  & 2  & 1  & 0  & 0  & 0  & -3  & 1  & -1  & -3  & -2  & 8 \\
\hline
2  & -2  & -2  & 2  & 1  & 2  & -2  & -1  & 2  & 6  & -1  & 1  & -3  & -1  & -5  & -10  & -2  & 9 \\
\hline
1  & -3  & 0  & 1  & 5  & 2  & -2  & 1  & -3  & 2  & 1  & 3  & -5  & -3  & -4  & -13  & 0  & 13 \\
\hline
\end{tabular}
} 
\caption{A sample of ten solutions found using the solution-mining method.}
\label{table:allanach_2}
\end{table}
The first of these solutions is equivalent to one found previously in Table~\ref{table:allanach_1}. This is because in the first anneal run the algorithm looks for solutions centered around zero, i.e. with entries between $[-1,2]$. Then it starts exploring the neighborhood of the solution found in the previous anneal run, gradually finding solutions with larger entries.

\section{\label{Sec:Conclusions}Conclusions}

 Diophantine problems in physics and beyond are often computationally hard. In this paper we have investigated the use of Ising machines for finding solutions to such problems, and have shown that they can succeed within search spaces that are vast. For example finding the (3,8,8) number 50139, the lowest number we find that can be written as the sum of eight cubes in two different ways, represents a search in a space of  size   $ 10^{24}$. Furthermore using ``solution mining'' for the task of anomaly cancellation, we find a proven ability to find solutions in search spaces of order $10^{26}$. 
 
 The methods described here are valid for any Ising machine, including both quantum and simulated annealers. We found that currently available quantum annealers can already solve a large variety of Diophantine problems at a relatively advanced level, but they are not yet competitive with their simulated counterparts. Nevertheless there are several reasons to believe that ultimately quantum annealers will become dominant for such tasks. The first is to do with the current obstacle to performing higher order tasks on a quantum annealer which is the fact that with the increasing complexity of the equations, finding a suitable embedding (i.e. an embedding such that the connectivity and the number of auxilliary qubits required are within the limits imposed by the quantum processing unit) is a non trivial task. Indeed it is worth noting that deciding whether a graph $G$ can be embedded in a graph $H$
 %with $G$ and $H$ arbitrary, is NP-hard. The complexity reduces to polynomial when $H$ is fixed \textit{a priori} \cite{robertson1995graph}. By contrast it 
 is itself an NP-complete problem (when $H$ is arbitrary but as in current annealers $G$ is built out of  \textit{Chimera} or \textit{Pegasus} sub-graphs) \cite{lobe2021minor}. 
 Thus one expects significant improvement in embedding as the overall size of the annealer, and the connectivity of its sub-graphs is continually increased.
 
 The second reason to be optimistic about quantum annealers is in the potential speed-up in the way that they find global minima. For example there are many techniques open to quantum annealers such as {\it diabatic} annealing that have the potential to avoid anneal times increasing exponentially with the difficulty of the problem~\cite{diabatic}, an issue that is seen in both {\it adiabatic} quantum annealing and in simulated annealing. These physical aspects of quantum annealing which were also crucial in the quantum field theory tunneling studies in Refs.~\cite{Abel:2020ebj, Abel:2020qzm, Abel:2021fpn} cannot be efficiently simulated classically. Indeed discrete problems generally favour quantum annealers because in a sense these machines can operate by performing a quantum gradient descent by tunnelling. By contrast any simulation method requires a defined dynamical process for its evolution, in order to hop between potential solutions, and this tends to become increasingly delicate with the difficulty of the problem. Thus for certain problems quantum tunneling could be an enormous advantage. For example, it has already been shown that quantum annealing overcomes simulated annealing in a large variety of cases (see for example Refs.~\cite{santoro2002theory,santoro2004,denchev2016computational}). Particularly interesting are the results found in Ref.~\cite{denchev2016computational} showing quantum annealing outperforming classical annealing by a factor of $10^8$ in finding the minimum of a particular crafted problem with tall and narrow energy barriers separating local minima. This is precisely the kind of configuration one expects when solving Diophantine problems. 
 
 In summary, we believe that the methods presented here should be an effective heuristic search method for the many discrete problems one encounters in physics. In particular it will be interesting to employ them in the string theory landscape context, and compare them to the genetic algorithmic and machine learning techniques that have been studied in Refs.~\cite{Abel:2014xta,He:2017,Ruehle:2017mzq,Carifio:2017bov, Bull:2018,Mutter:2018sra,Klaewer:2018sfl,Brodie:2019dfx,Bull:2019cij,Cole:2019enn,Halverson:2019tkf,Ruehle:2020jrk,Krippendorf:2021uxu, AbdusSalam:2020ywo,Bena:2020xrh,Larfors:2020ugo,Constantin:2021for,Abel:2021rrj,Loges:2021hvn}. \\

\noindent {\it {Acknowledgements}:}  We would like to thank Ben Allanach,  Andrei Constantin, Juan Criado, Jim Halverson, Thomas Harvey, Andre Lukas, Fabian Ruehle and Michael Spannowsky  for helpful discussions. S.A. is supported by the STFC under grant ST/P001246/1. 

\bibliographystyle{inspire}
\bibliography{references,referencesSAMS}

\providecommand{\href}[2]{#2}\begingroup\raggedright\begin{thebibliography}{10}

\bibitem{tHooft:1979rat}
G.~'t~Hooft, ``{Naturalness, chiral symmetry, and spontaneous chiral symmetry
  breaking},'' \href{http://dx.doi.org/10.1007/978-1-4684-7571-5_9}{NATO Sci.
  Ser. B {\bfseries 59} (1980) 135--157}.

\bibitem{Allanach:2018vjg}
B.~C. Allanach, J.~Davighi, and S.~Melville, ``{An Anomaly-free Atlas: charting
  the space of flavour-dependent gauged $U(1)$ extensions of the Standard
  Model},'' \href{http://dx.doi.org/10.1007/JHEP02(2019)082}{JHEP {\bfseries
  02} (2019) 082}, \href{http://arxiv.org/abs/1812.04602}{[arXiv:1812.04602
  [hep-ph]]}. [Erratum: JHEP 08, 064 (2019)].

\bibitem{Allanach:2019uuu}
B.~C. Allanach, B.~Gripaios, and J.~Tooby-Smith, ``{Solving local anomaly
  equations in gauge-rank extensions of the Standard Model},''
  \href{http://dx.doi.org/10.1103/PhysRevD.101.075015}{Phys. Rev. D {\bfseries
  101} no.~7, (2020) 075015},
  \href{http://arxiv.org/abs/1912.10022}{[arXiv:1912.10022 [hep-th]]}.

\bibitem{Allanach:2020zna}
B.~C. Allanach, B.~Gripaios, and J.~Tooby-Smith, ``{Anomaly cancellation with
  an extra gauge boson},''
  \href{http://dx.doi.org/10.1103/PhysRevLett.125.161601}{Phys. Rev. Lett.
  {\bfseries 125} no.~16, (2020) 161601},
  \href{http://arxiv.org/abs/2006.03588}{[arXiv:2006.03588 [hep-th]]}.

\bibitem{Allanach:2021bfe}
B.~C. Allanach, B.~Gripaios, and J.~Tooby-Smith, ``{Semisimple extensions of
  the Standard Model gauge algebra},''
  \href{http://dx.doi.org/10.1103/PhysRevD.104.035035}{Phys. Rev. D {\bfseries
  104} no.~3, (2021) 035035},
  \href{http://arxiv.org/abs/2104.14555}{[arXiv:2104.14555 [hep-th]]}.

\bibitem{TUNG1987324}
S.~Tung, ``Computational complexities of diophantine equations with
  parameters,''
  \href{http://dx.doi.org/https://doi.org/10.1016/0196-6774(87)90013-7}{Journal
  of Algorithms {\bfseries 8} no.~3, (1987) 324--336}.
  \url{https://www.sciencedirect.com/science/article/pii/0196677487900137}.

\bibitem{Denef:2006ad}
F.~Denef and M.~R. Douglas, ``{Computational complexity of the landscape.
  I.},'' \href{http://dx.doi.org/10.1016/j.aop.2006.07.013}{Annals Phys.
  {\bfseries 322} (2007) 1096--1142},
  \href{http://arxiv.org/abs/hep-th/0602072}{[arXiv:hep-th/0602072]}.

\bibitem{Denef:2017cxt}
F.~Denef, M.~R. Douglas, B.~Greene, and C.~Zukowski, ``{Computational
  complexity of the landscape II\textemdash{}Cosmological considerations},''
  \href{http://dx.doi.org/10.1016/j.aop.2018.03.013}{Annals Phys. {\bfseries
  392} (2018) 93--127},
  \href{http://arxiv.org/abs/1706.06430}{[arXiv:1706.06430 [hep-th]]}.

\bibitem{Halverson:2018cio}
J.~Halverson and F.~Ruehle, ``{Computational Complexity of Vacua and Near-Vacua
  in Field and String Theory},''
  \href{http://dx.doi.org/10.1103/PhysRevD.99.046015}{Phys. Rev. D {\bfseries
  99} no.~4, (2019) 046015},
  \href{http://arxiv.org/abs/1809.08279}{[arXiv:1809.08279 [hep-th]]}.

\bibitem{Halverson:2019vmd}
J.~Halverson, M.~Plesser, F.~Ruehle, and J.~Tian, ``{K\"ahler Moduli
  Stabilization and the Propagation of Decidability},''
  \href{http://dx.doi.org/10.1103/PhysRevD.101.046010}{Phys. Rev. D {\bfseries
  101} no.~4, (2020) 046010},
  \href{http://arxiv.org/abs/1911.07835}{[arXiv:1911.07835 [hep-th]]}.

\bibitem{PhysRevE.58.5355}
T.~Kadowaki and H.~Nishimori, ``Quantum annealing in the transverse ising
  model,'' \href{http://dx.doi.org/10.1103/PhysRevE.58.5355}{Phys. Rev. E
  {\bfseries 58} (Nov, 1998) 5355--5363}.
  \url{https://link.aps.org/doi/10.1103/PhysRevE.58.5355}.

\bibitem{doi:10.1126/science.1057726}
E.~Farhi, J.~Goldstone, S.~Gutmann, J.~Lapan, A.~Lundgren, and D.~Preda, ``A
  quantum adiabatic evolution algorithm applied to random instances of an
  np-complete problem,''
  \href{http://dx.doi.org/10.1126/science.1057726}{Science {\bfseries 292}
  no.~5516, (2001) 472--475},
  \href{http://arxiv.org/abs/https://www.science.org/doi/pdf/10.1126/science.1057726}{[https://www.science.org/doi/pdf/10.1126/science.1057726]}.
  \url{https://www.science.org/doi/abs/10.1126/science.1057726}.

\bibitem{RevModPhys.80.1061}
A.~Das and B.~K. Chakrabarti, ``Colloquium: Quantum annealing and analog
  quantum computation,''
  \href{http://dx.doi.org/10.1103/RevModPhys.80.1061}{Rev. Mod. Phys.
  {\bfseries 80} (Sep, 2008) 1061--1081}.
  \url{https://link.aps.org/doi/10.1103/RevModPhys.80.1061}.

\bibitem{10.3389/fphy.2014.00005}
A.~Lucas, ``Ising formulations of many np problems,''
  \href{http://dx.doi.org/10.3389/fphy.2014.00005}{Frontiers in Physics
  {\bfseries 2} (2014) }.
  \url{https://www.frontiersin.org/article/10.3389/fphy.2014.00005}.

\bibitem{Hauke_2020}
P.~Hauke, H.~G. Katzgraber, W.~Lechner, H.~Nishimori, and W.~D. Oliver,
  ``Perspectives of quantum annealing: methods and implementations,''
  \href{http://dx.doi.org/10.1088/1361-6633/ab85b8}{Reports on Progress in
  Physics {\bfseries 83} no.~5, (May, 2020) 054401}.
  \url{https://doi.org/10.1088/1361-6633/ab85b8}.

\bibitem{Mohseni}
N.~Mohseni, P.~McMahon, and T.~Byrnes, ``Ising machines as hardware solvers of
  combinatorial optimization problems.'' Nature Reviews Physics (2022) .
  \url{https://doi.org/10.1038/s42254-022-00440-8}.

\bibitem{Guy:1994a}
R.~K. Guy, ``Sums of like powers. euler's conjecture, §d1 in unsolved problems
  in number theory,'' 1994.

\bibitem{Meyrignac}
J.-C. Meyrignac, ``Computing minimal equal sums of like powers.''
\newblock \url{http://euler.free.fr}.

\bibitem{Eulernet}
Eulernet.
\newblock \url{http://euler.free.fr/database.txt}.

\bibitem{Weisstein}
E.~W. Weisstein, ``Diophantine equation--nth powers (replace ``n'' by the
  power): From mathworld--a wolfram web resource.''
\newblock
  \url{https://mathworld.wolfram.com/DiophantineEquationnthPowers.html}.

\bibitem{piezas}
T.~Piezas, ``A collection of algebraic identities.''
\newblock \url{https://sites.google.com/site/tpiezas/Home}.

\bibitem{Gadde:2013lxa}
A.~Gadde, S.~Gukov, and P.~Putrov, ``{(0, 2) trialities},''
  \href{http://dx.doi.org/10.1007/JHEP03(2014)076}{JHEP {\bfseries 03} (2014)
  076}, \href{http://arxiv.org/abs/1310.0818}{[arXiv:1310.0818 [hep-th]]}.

\bibitem{Franco:2016nwv}
S.~Franco, S.~Lee, and R.-K. Seong, ``{Brane brick models and 2d (0, 2)
  triality},'' \href{http://dx.doi.org/10.1007/JHEP05(2016)020}{JHEP {\bfseries
  05} (2016) 020}, \href{http://arxiv.org/abs/1602.01834}{[arXiv:1602.01834
  [hep-th]]}.

\bibitem{Franco:2016tcm}
S.~Franco, S.~Lee, R.-K. Seong, and C.~Vafa, ``{Quadrality for Supersymmetric
  Matrix Models},'' \href{http://dx.doi.org/10.1007/JHEP07(2017)053}{JHEP
  {\bfseries 07} (2017) 053},
  \href{http://arxiv.org/abs/1612.06859}{[arXiv:1612.06859 [hep-th]]}.

\bibitem{dattani2014quantum}
N.~S. Dattani and N.~Bryans, ``Quantum factorization of 56153 with only 4
  qubits,'' arXiv preprint (2014) . \url{https://arxiv.org/abs/1411.6758}.

\bibitem{tanburn:2015a}
R.~Tanburn, E.~Okada, and N.~Dattani, ``Reducing multi-qubit interactions in
  adiabatic quantum computation without adding auxiliary qubits. part 1: The
  "deduc-reduc" method and its application to quantum factorization of
  numbers,'' 2015.
\newblock \url{https://arxiv.org/abs/1508.04816}.

\bibitem{Jiang:2018b}
S.~Jiang, K.~A. Britt, A.~J. McCaskey, T.~S. Humble, and S.~Kais, ``Quantum
  annealing for prime factorization,'' 2018.
\newblock \url{https://arxiv.org/abs/1804.02733}.

\bibitem{peng2019factoring}
W.~Peng, B.~Wang, F.~Hu, Y.~Wang, X.~Fang, X.~Chen, and C.~Wang, ``Factoring
  larger integers with fewer qubits via quantum annealing with optimized
  parameters,'' SCIENCE CHINA Physics, Mechanics \& Astronomy {\bfseries 62}
  no.~6, (2019) 1--8. \url{https://doi.org/10.1007/s11433-018-9307-1}.

\bibitem{warren2019factoring}
R.~H. Warren, ``Factoring on a quantum annealing computer,'' Quantum
  Information \& Computation {\bfseries 19} no.~3-4, (2019) 252--261.

\bibitem{wang2020prime}
B.~Wang, F.~Hu, H.~Yao, and C.~Wang, ``Prime factorization algorithm based on
  parameter optimization of {I}sing model,'' Scientific reports {\bfseries 10}
  no.~1, (2020) 1--10.
  \url{https://www.nature.com/articles/s41598-020-62802-5}.

\bibitem{chang2019quantum}
C.~C. Chang, A.~Gambhir, T.~S. Humble, and S.~Sota, ``Quantum annealing for
  systems of polynomial equations,''
  \href{http://dx.doi.org/10.1038/s41598-019-46729-0}{Scientific Reports
  {\bfseries 9} no.~1, (Jul, 2019) }.
  \url{https://doi.org/10.1038%2Fs41598-019-46729-0}.

\bibitem{chang2019least}
T.~H. Chang, T.~C. Lux, and S.~S. Tipirneni, ``Least-squares solutions to
  polynomial systems of equations with quantum annealing,'' Quantum Information
  Processing {\bfseries 18} no.~12, (2019) 1--17.
  \url{https://doi.org/10.1007/s11128-019-2489-x}.

\bibitem{rosenberg1975reduction}
I.~G. Rosenberg, ``Reduction of bivalent maximization to the quadratic case.''
  Cahiers du Centre d’Etudes de Recherche Operationnelle {\bfseries 17}
  (1975) 71–74.

\bibitem{dattani:2019a}
N.~Dattani, ``Quadratization in discrete optimization and quantum mechanics,''
  2019.
\newblock \url{https://arxiv.org/abs/1901.04405}.

\bibitem{Abel:2022lqr}
S.~Abel, J.~C. Criado, and M.~Spannowsky, ``{Completely Quantum Neural
  Networks},'' \href{http://arxiv.org/abs/2202.11727}{[arXiv:2202.11727
  [quant-ph]]}.

\bibitem{babbush2013resource}
R.~Babbush, B.~O{\textquotesingle}Gorman, and A.~Aspuru-Guzik, ``Resource
  efficient gadgets for compiling adiabatic quantum optimization problems,''
  \href{http://dx.doi.org/10.1002/andp.201300120}{Annalen der Physik {\bfseries
  525} no.~10-11, (Sep, 2013) 877--888}.
  \url{https://doi.org/10.1002%2Fandp.201300120}.

\bibitem{boros2002pseudo}
E.~Boros and P.~L. Hammer, ``Pseudo-{B}oolean optimization,'' Discrete Applied
  Mathematics {\bfseries 123} no.~1-3, (2002) 155--225.
  \url{https://doi.org/10.1016/S0166-218X(01)00341-9}.

\bibitem{chvatal1979greedy}
V.~Chvatal, ``A greedy heuristic for the set-covering problem,'' Mathematics of
  operations research {\bfseries 4} no.~3, (1979) 233--235.
  \url{https://doi.org/10.1287/moor.4.3.233}.

\bibitem{Aguiar2014DiophantineEA}
H.~Aguiar, ``Diophantine {E}quations and {F}uzzy {A}daptive {S}imulated
  {A}nnealing,''
\newblock 2014.

\bibitem{moradi2019solving}
M.~Moradi, ``Solving ill-conditioned linear equations using {S}imulated
  {A}nnealing method,'' Journal of Hyperstructures {\bfseries 7} no.~1, (2019)
  . \url{http://www.jhs-uma.com/index.php/JHSMS/article/viewFile/327/120}.

\bibitem{abraham2001diophantine}
S.~Abraham and M.~Sanglikar, ``A {D}iophantine equation solver-a genetic
  algorithm application,'' in {\em Mathematical Colloquium journal}, vol.~15.
\newblock 2001.

\bibitem{abraham2010particle}
S.~Abraham, S.~Sanyal, and M.~Sanglikar, ``Particle swarm optimization based
  diophantine equation solver,'' 2010.
\newblock \url{https://arxiv.org/abs/1003.2724}.

\bibitem{abraham2013finding}
S.~Abraham, S.~Sanyal, and M.~Sanglikar, ``Finding numerical solutions of
  {D}iophantine equations using ant colony optimization,'' Applied Mathematics
  and Computation {\bfseries 219} no.~24, (2013) 11376--11387.
  \url{https://doi.org/10.1016/j.amc.2013.05.051}.

\bibitem{perez2013numerical}
O.~P{\'e}rez, I.~Amaya, and R.~Correa, ``Numerical solution of certain
  exponential and non-linear {D}iophantine systems of equations by using a
  discrete particle swarm optimization algorithm,'' Applied Mathematics and
  Computation {\bfseries 225} (2013) 737--746.
  \url{https://doi.org/10.1016/j.amc.2013.10.007}.

\bibitem{LantingAQC2017}
T.~Lanting, ``The {D-Wave 2000Q Processor},'' 2017.
\newblock
  \url{https://www.youtube.com/watch?v=_Y9sVY-XBfI&index=2&list=PLAlerseOylzZw8A-R4lyU3mqfYfQSggiz}.
  presented at AQC 2017.

\bibitem{Boyer}
C.~Boyer, ``New {U}pper {B}ounds for {T}axicab and {C}abtaxi {N}umbers.''
\newblock \url{http://www.christianboyer.com/taxicab/}.

\bibitem{Langacker:2009}
P.~Langacker, ``The physics of heavy ${Z}^{\ensuremath{'}}$ gauge bosons,''
  \href{http://dx.doi.org/10.1103/RevModPhys.81.1199}{Rev. Mod. Phys.
  {\bfseries 81} (Aug, 2009) 1199--1228}.
  \url{https://link.aps.org/doi/10.1103/RevModPhys.81.1199}.

\bibitem{zyla2020}
P.~Z. \textit{et al.} (Particle Data~Group), ``{Review of Particle Physics},''
  \href{http://dx.doi.org/10.1093/ptep/ptaa104}{Progress of Theoretical and
  Experimental Physics {\bfseries 2020} no.~8, (08, 2020) },
  \href{http://arxiv.org/abs/https://academic.oup.com/ptep/article-pdf/2020/8/083C01/34673722/ptaa104.pdf}{[https://academic.oup.com/ptep/article-pdf/2020/8/083C01/34673722/ptaa104.pdf]}.
  \url{https://doi.org/10.1093/ptep/ptaa104}. 083C01.

\bibitem{Appelquist:2003}
T.~Appelquist, B.~A. Dobrescu, and A.~R. Hopper, ``Nonexotic neutral gauge
  bosons,'' \href{http://dx.doi.org/10.1103/PhysRevD.68.035012}{Phys. Rev. D
  {\bfseries 68} (Aug, 2003) 035012}.
  \url{https://link.aps.org/doi/10.1103/PhysRevD.68.035012}.

\bibitem{Batra:2005rh}
P.~Batra, B.~A. Dobrescu, and D.~Spivak, ``{Anomaly-free sets of fermions},''
  \href{http://dx.doi.org/10.1063/1.2222081}{J. Math. Phys. {\bfseries 47}
  (2006) 082301},
  \href{http://arxiv.org/abs/hep-ph/0510181}{[arXiv:hep-ph/0510181]}.

\bibitem{liu2012searching}
J.-Y. Liu, Y.~Tang, and Y.-L. Wu, ``{Searching for $Z^{'}$ Gauge Boson in an
  Anomaly-Free U(1)$'$ Gauge Family Model},''
  \href{http://dx.doi.org/10.1088/0954-3899/39/5/055003}{J. Phys. G {\bfseries
  39} (2012) 055003}, \href{http://arxiv.org/abs/1108.5012}{[arXiv:1108.5012
  [hep-ph]]}.

\bibitem{ismail2017axial}
A.~Ismail, W.-Y. Keung, K.-H. Tsao, and J.~Unwin, ``{Axial vector $Z'$ and
  anomaly cancellation},''
  \href{http://dx.doi.org/10.1016/j.nuclphysb.2017.03.001}{Nucl. Phys. B
  {\bfseries 918} (2017) 220--244},
  \href{http://arxiv.org/abs/1609.02188}{[arXiv:1609.02188 [hep-ph]]}.

\bibitem{tang2018flavor}
Y.~Tang and Y.-L. Wu, ``{Flavor non-universal gauge interactions and anomalies
  in {B}-meson decays},''
  \href{http://dx.doi.org/10.1088/1674-1137/42/3/033104}{Chin. Phys. C
  {\bfseries 42} no.~3, (2018) 033104},
  \href{http://arxiv.org/abs/1705.05643}{[arXiv:1705.05643 [hep-ph]]}.
  [Erratum: Chin.Phys.C 44, 069101 (2020)].

\bibitem{Ellis:2018}
J.~Ellis, M.~Fairbairn, and P.~Tunney, ``Anomaly-free models for flavour
  anomalies,''
  \href{http://dx.doi.org/https://doi.org/10.1140/epjc/s10052-018-5725-0}{The
  European Physical Journal C {\bfseries 78} no.~3, (2018) 1--9}.

\bibitem{Costa:2019anomaly}
D.~B. Costa, B.~A. Dobrescu, and P.~J. Fox, ``General solution to the ${U}(1)$
  anomaly equations,''
  \href{http://dx.doi.org/10.1103/PhysRevLett.123.151601}{Phys. Rev. Lett.
  {\bfseries 123} (Oct, 2019) 151601}.
  \url{https://link.aps.org/doi/10.1103/PhysRevLett.123.151601}.

\bibitem{Smolkovivc2019anomaly}
A.~Smolkovi{\v{c}}, M.~Tammaro, and J.~Zupan, ``Anomaly free
  {F}roggatt-{N}ielsen models of flavor,''
  \href{http://dx.doi.org/https://doi.org/10.1007/JHEP10(2019)188}{Journal of
  High Energy Physics {\bfseries 2019} no.~10, (2019) 1--50}.

\bibitem{Allanach:2020geometric}
B.~C. Allanach, B.~Gripaios, and J.~Tooby-Smith, ``{Geometric General Solution
  to the $U(1)$ Anomaly Equations},''
  \href{http://dx.doi.org/10.1007/JHEP05(2020)065}{JHEP {\bfseries 05} (2020)
  065}, \href{http://arxiv.org/abs/1912.04804}{[arXiv:1912.04804 [hep-th]]}.

\bibitem{Costa:2020}
D.~B. Costa, ``Anomaly-free ${U(1)}^{m}$ extensions of the {S}tandard
  {M}odel,'' \href{http://dx.doi.org/10.1103/PhysRevD.102.115006}{Phys. Rev. D
  {\bfseries 102} (Dec, 2020) 115006}.
  \url{https://link.aps.org/doi/10.1103/PhysRevD.102.115006}.

\bibitem{Podo:2022gyj}
A.~Podo and F.~Revello, ``{Integer solutions to the anomaly equations for a
  class of chiral gauge theories},''
  \href{http://arxiv.org/abs/2205.03428}{[arXiv:2205.03428 [hep-th]]}.

\bibitem{lobe2021minor}
E.~Lobe and A.~Lutz, ``Minor embedding in broken chimera and pegasus graphs is
  np-complete,'' 2021.
\newblock \url{https://arxiv.org/abs/2110.08325}.

\bibitem{diabatic}
E.~Crosson and D.~Lidar, ``Prospects for quantum enhancement with diabatic
  quantum annealing,'' \href{http://dx.doi.org/10.1038/s42254-021-00313-6}{Nat.
  Rev. Phys. {\bfseries 3} (Sep, 2021) 466–--489}.
  \url{https://doi.org/10.1038/s42254-021-00313-6}.

\bibitem{Abel:2020ebj}
S.~Abel, N.~Chancellor, and M.~Spannowsky, ``{Quantum Computing for Quantum
  Tunnelling},'' \href{http://arxiv.org/abs/2003.07374}{[arXiv:2003.07374
  [hep-ph]]}.

\bibitem{Abel:2020qzm}
S.~Abel and M.~Spannowsky, ``{Observing the fate of the false vacuum with a
  quantum laboratory},''
  \href{http://dx.doi.org/10.1103/PRXQuantum.2.010349}{P. R. X. Quantum.
  {\bfseries 2} (2021) 010349},
  \href{http://arxiv.org/abs/2006.06003}{[arXiv:2006.06003 [hep-th]]}.

\bibitem{Abel:2021fpn}
S.~Abel, A.~Blance, and M.~Spannowsky, ``{Quantum Optimisation of Complex
  Systems with a Quantum Annealer},''
  \href{http://arxiv.org/abs/2105.13945}{[arXiv:2105.13945 [quant-ph]]}.

\bibitem{santoro2002theory}
G.~E. Santoro, R.~Marton{\'a}k, E.~Tosatti, and R.~Car, ``Theory of {Q}uantum
  {A}nnealing of an {I}sing {S}pin {G}lass,''
  \href{http://dx.doi.org/10.1126/science.1068774}{Science {\bfseries 295}
  no.~5564, (Mar, 2002) 2427--2430}.
  \url{https://doi.org/10.1126%2Fscience.1068774}.

\bibitem{santoro2004}
R.~Marto\ifmmode~\check{n}\else \v{n}\fi{}\'ak, G.~E. Santoro, and E.~Tosatti,
  ``Quantum annealing of the traveling-salesman problem,''
  \href{http://dx.doi.org/10.1103/PhysRevE.70.057701}{Phys. Rev. E {\bfseries
  70} (Nov, 2004) 057701}.
  \url{https://link.aps.org/doi/10.1103/PhysRevE.70.057701}.

\bibitem{denchev2016computational}
V.~S. Denchev, S.~Boixo, S.~V. Isakov, N.~Ding, R.~Babbush, V.~Smelyanskiy,
  J.~Martinis, and H.~Neven, ``What is the computational value of finite-range
  tunneling?'' Physical Review X {\bfseries 6} no.~3, (2016) 031015.
  \url{https://journals.aps.org/prx/abstract/10.1103/PhysRevX.6.031015}.

\bibitem{Abel:2014xta}
S.~Abel and J.~Rizos, ``{Genetic {A}lgorithms and the {S}earch for {V}iable
  {S}tring {V}acua},'' \href{http://dx.doi.org/10.1007/JHEP08(2014)010}{JHEP
  {\bfseries 08} (2014) 010},
  \href{http://arxiv.org/abs/1404.7359}{[arXiv:1404.7359 [hep-th]]}.

\bibitem{He:2017}
Y.-H. He, ``Machine-learning the string landscape,''
  \href{http://dx.doi.org/https://doi.org/10.1016/j.physletb.2017.10.024}{Physics
  Letters B {\bfseries 774} (2017) 564--568}.
  \url{https://www.sciencedirect.com/science/article/pii/S0370269317308365}.

\bibitem{Ruehle:2017mzq}
F.~Ruehle, ``{Evolving neural networks with genetic algorithms to study the
  String Landscape},'' \href{http://dx.doi.org/10.1007/JHEP08(2017)038}{JHEP
  {\bfseries 08} (2017) 038},
  \href{http://arxiv.org/abs/1706.07024}{[arXiv:1706.07024 [hep-th]]}.

\bibitem{Carifio:2017bov}
J.~Carifio, J.~Halverson, D.~Krioukov, and B.~D. Nelson, ``{Machine Learning in
  the String Landscape},''
  \href{http://dx.doi.org/10.1007/JHEP09(2017)157}{JHEP {\bfseries 09} (2017)
  157}, \href{http://arxiv.org/abs/1707.00655}{[arXiv:1707.00655 [hep-th]]}.

\bibitem{Bull:2018}
K.~Bull, Y.-H. He, V.~Jejjala, and C.~Mishra, ``Machine learning {CICY}
  threefolds,''
  \href{http://dx.doi.org/https://doi.org/10.1016/j.physletb.2018.08.008}{Physics
  Letters B {\bfseries 785} (2018) 65--72}.
  \url{https://www.sciencedirect.com/science/article/pii/S0370269318306117}.

\bibitem{Mutter:2018sra}
A.~M\"utter, E.~Parr, and P.~K.~S. Vaudrevange, ``{Deep learning in the
  heterotic orbifold landscape},''
  \href{http://dx.doi.org/10.1016/j.nuclphysb.2019.01.013}{Nucl. Phys. B
  {\bfseries 940} (2019) 113--129},
  \href{http://arxiv.org/abs/1811.05993}{[arXiv:1811.05993 [hep-th]]}.

\bibitem{Klaewer:2018sfl}
D.~Klaewer and L.~Schlechter, ``{Machine Learning Line Bundle Cohomologies of
  Hypersurfaces in Toric Varieties},''
  \href{http://dx.doi.org/10.1016/j.physletb.2019.01.002}{Phys. Lett. B
  {\bfseries 789} (2019) 438--443},
  \href{http://arxiv.org/abs/1809.02547}{[arXiv:1809.02547 [hep-th]]}.

\bibitem{Brodie:2019dfx}
C.~R. Brodie, A.~Constantin, R.~Deen, and A.~Lukas, ``{Machine Learning Line
  Bundle Cohomology},''
  \href{http://dx.doi.org/10.1002/prop.201900087}{Fortsch. Phys. {\bfseries 68}
  no.~1, (2020) 1900087},
  \href{http://arxiv.org/abs/1906.08730}{[arXiv:1906.08730 [hep-th]]}.

\bibitem{Bull:2019cij}
K.~Bull, Y.-H. He, V.~Jejjala, and C.~Mishra, ``{Getting CICY High},''
  \href{http://dx.doi.org/10.1016/j.physletb.2019.06.067}{Phys. Lett. B
  {\bfseries 795} (2019) 700--706},
  \href{http://arxiv.org/abs/1903.03113}{[arXiv:1903.03113 [hep-th]]}.

\bibitem{Cole:2019enn}
A.~Cole, A.~Schachner, and G.~Shiu, ``{Searching the Landscape of Flux Vacua
  with Genetic Algorithms},''
  \href{http://dx.doi.org/10.1007/JHEP11(2019)045}{JHEP {\bfseries 11} (2019)
  045}, \href{http://arxiv.org/abs/1907.10072}{[arXiv:1907.10072 [hep-th]]}.

\bibitem{Halverson:2019tkf}
J.~Halverson, B.~Nelson, and F.~Ruehle, ``{Branes with Brains: Exploring String
  Vacua with Deep Reinforcement Learning},''
  \href{http://dx.doi.org/10.1007/JHEP06(2019)003}{JHEP {\bfseries 06} (2019)
  003}, \href{http://arxiv.org/abs/1903.11616}{[arXiv:1903.11616 [hep-th]]}.

\bibitem{Ruehle:2020jrk}
F.~Ruehle, ``{Data science applications to string theory},''
  \href{http://dx.doi.org/10.1016/j.physrep.2019.09.005}{Phys. Rept. {\bfseries
  839} (2020) 1--117}.

\bibitem{Krippendorf:2021uxu}
S.~Krippendorf, R.~Kroepsch, and M.~Syvaeri, ``{Revealing systematics in
  phenomenologically viable flux vacua with reinforcement learning},''
  \href{http://arxiv.org/abs/2107.04039}{[arXiv:2107.04039 [hep-th]]}.

\bibitem{AbdusSalam:2020ywo}
S.~AbdusSalam, S.~Abel, M.~Cicoli, F.~Quevedo, and P.~Shukla, ``{A systematic
  approach to K\"ahler moduli stabilisation},''
  \href{http://dx.doi.org/10.1007/JHEP08(2020)047}{JHEP {\bfseries 08} no.~08,
  (2020) 047}, \href{http://arxiv.org/abs/2005.11329}{[arXiv:2005.11329
  [hep-th]]}.

\bibitem{Bena:2020xrh}
I.~Bena, J.~Bl\r{a}b\"ack, M.~Gra\~na, and S.~L\"ust, ``{The tadpole
  problem},'' \href{http://dx.doi.org/10.1007/JHEP11(2021)223}{JHEP {\bfseries
  11} (2021) 223}, \href{http://arxiv.org/abs/2010.10519}{[arXiv:2010.10519
  [hep-th]]}.

\bibitem{Larfors:2020ugo}
M.~Larfors and R.~Schneider, ``{Explore and Exploit with Heterotic Line Bundle
  Models},'' \href{http://dx.doi.org/10.1002/prop.202000034}{Fortsch. Phys.
  {\bfseries 68} no.~5, (2020) 2000034},
  \href{http://arxiv.org/abs/2003.04817}{[arXiv:2003.04817 [hep-th]]}.

\bibitem{Constantin:2021for}
A.~Constantin, T.~R. Harvey, and A.~Lukas, ``{Heterotic {S}tring {M}odel
  {B}uilding with {M}onad {B}undles and {R}einforcement {L}earning},''
  \href{http://arxiv.org/abs/2108.07316}{[arXiv:2108.07316 [hep-th]]}.

\bibitem{Abel:2021rrj}
S.~Abel, A.~Constantin, T.~R. Harvey, and A.~Lukas, ``{Evolving Heterotic Gauge
  Backgrounds: Genetic Algorithms versus Reinforcement Learning},''
  \href{http://dx.doi.org/10.1002/prop.202200034}{Fortsch. Phys. {\bfseries 70}
  no.~5, (2022) 2200034},
  \href{http://arxiv.org/abs/2110.14029}{[arXiv:2110.14029 [hep-th]]}.

\bibitem{Loges:2021hvn}
G.~J. Loges and G.~Shiu, ``{Breeding Realistic D-Brane Models},''
  \href{http://dx.doi.org/10.1002/prop.202200038}{Fortsch. Phys. {\bfseries 70}
  no.~5, (2022) 2200038},
  \href{http://arxiv.org/abs/2112.08391}{[arXiv:2112.08391 [hep-th]]}.

\end{thebibliography}\endgroup

\end{document}